\documentclass[a4paper,11pt]{article}
\pdfoutput=1 % if your are submitting a pdflatex (i.e. if you have
\usepackage{jinstpub} % for details on the use of the package, please
\title{\boldmath Exploration of optimized front-end readout circuit for time measurement of large-area SiPM arrays}

\author[a]{M.X. Wang,}
\author[a]{Y. Liu,}
\author[a]{Y.Q. Tan,}
\author[a]{J.N. Tang,}
\author[a,1]{W.H. Wu\note{Corresponding author.},}
\author[b,a]{D.L. Xu,}
\author[a]{W. Zhi}
\author[a]{and Z.Z. Zhou}

\affiliation[a]{School of Physics and Astronomy, Shanghai Jiao Tong University\\800 Dongchuan Road, Minhang District, Shanghai, China}
\affiliation[b]{Tsung-Dao Lee Institute, Shanghai Jiao Tong Univeristy\\1 Lisuo Road, Pudong New Area, Shanghai, China}

\emailAdd{wuweihao@sjtu.edu.cn}

\abstract{The detector of TRopIcal DEep-sea Neutrino Telescope (TRIDENT) will use large-area silicon photomultiplier (SiPM) arrays combined with photomultiplier tubes to boost photon detection efficiency and pointing capability. An application-specific integrated circuit (ASIC) is being developed to aim at high-resolution time measurement of large-area SiPM arrays. This work researches four architectures of readout circuits including different input stages (common gate stage and negative feedback common gate stage) and discriminators (two types of current discriminator and one voltage discriminator) using a 180 nm CMOS process for optimizing time resolution. The experimental measurements show that single photon time resolutions performed using Hamamatsu S13360-3050PE SiPMs are around 260 ps full width at half maximum (FWHM). A timing jitter less than 500 ps FWHM when connecting a 6 $\times$ 6 mm$^{2}$ SiPM array is achieved. The power consumption is less than 7 mW/channel. Additionally, a digital summation is applied to reduce the number of output interfaces. The measured performances of the ASIC cater to the TRIDENT application requirements.}


\begin{document}
\maketitle
\flushbottom

\section{Introduction}
\label{sec:intro}
TRopIcal DEep-sea Neutrino Telescope (TRIDENT) is a next-generation multi-cubic-kilometre neutrino telescope planned to be constructed in the South China Sea \cite{Trident}. TRIDENT pathfinder experiment was completed in~2021 and found a site suitable for the construction of a neutrino telescope. Two independent light measurement systems were deployed using photomultiplier tubes (PMTs) and cameras to decode the optical properties of the seawater in situ \cite{TridentE}. All previous or current operational neutrino telescopes use photomultiplier tubes to detect Cherenkov light generated by the secondary particles from neutrino interactions, which has a typical transit time spread at a nanosecond level. The innovative design of TRIDENT hybrid digital optical modules (hDOM) utilizes silicon photomultipliers (SiPMs) in addition to PMTs to improve the detector’s photon coverage and time resolution. 

Each SiPM array has a detecting area of about 7~cm$^{2}$, which comprises 64 SiPMs with a size of 3 $\times$ 3~mm$^{2}$ \cite{hDOM}. Furthermore, the SiPM array needs to have the ability to detect single photons due to the small number of Cherenkov photons, which limits the amplitude of the input signal. Using large-area SiPM arrays as photodetectors in neutrino telescopes is not an easy task. The dark rate of the SiPM is at a level of 30~kHz/mm$^{2}$ \cite{sipmdcr}, which is much larger than that of the PMT at room temperature. However, the ambient temperature at the deep sea (2~$^\circ C$ $\sim$ 4~$^\circ C$) will help to reduce the dark rate by an order of magnitude. In addition, the trigger scheme via PMTs can further reduce the impact of the dark rate. When a leading-edge discriminator is used to determine the photon time-of-arrival, equation \eqref{eq:eq1} describes how the electronics jitter ($\sigma_{t}$), measured as the standard deviation of the time-of-arrival (ToA), is affected by the rise time and amplitude of the input signal, and the integrated electronic output noise ($\sigma_{n}$). 
\begin{equation}
	\label{eq:eq1}
	\begin{split}
		\sigma_{t} \approx \frac{RiseTime \times \sigma_{n}}{Amplitude}.
	\end{split}
\end{equation}
Increasing the bias voltage of SiPMs can improve the signal-to-noise ratio. However, this operation will increase the dark rate. The combined readout of a large area SiPM array implies a large parasitic capacitance, which severely limits the bandwidth of the front-end circuit and increases the rise time. In addition, the limited space and power budget are important to consider when designing front-end readout electronics for SiPMs.

We have explored the readout scheme using commercial discrete devices \cite{sipm1}. The measured single photon time resolution (SPTR) of a 3 $\times$ 3~mm$^{2}$ SiPM (Hamamatsu S13360-3050PE) is about 200 ps full width at half maximum (FWHM). 
Several SiPMs are connected in series and parallel to form a large-area array. The outputs of multiple channels are combined into one channel through an analogue summing circuit. The measured SPTR of a 4 $\times$ 4 SiPM array is about 300~ps FWHM with a power consumption of about 100~mW \cite{sipm2}. 

Compared to the aforementioned discrete device approach, using application-specific integrated circuits (ASIC) for readout has the advantages of higher integration and lower power consumption. Several ASICs have been developed by other institutes for SiPM readout to provide precise time measurements. For instance, NINO \cite{NINO} is an 8-channel ASIC that can be used for time measurements of Hamamatsu S13360-3050CS SiPMs. The measured SPTR is 160~ps FWHM \cite{HRFlexToT}. However, the power consumption of NINO is high (27~mW/channel). FlexToT \cite{FlexToT}, HRFlexToT \cite{HRFlexToT}, Petiroc2A \cite{Petiroc2A} and TOFPET2 \cite{TOFPET2} are alternative ASICs that provide both time and energy measurements. In hDOMs, SiPMs are expected to provide time information of photons only. Petiroc2A and TOFPET2 include internal time-to-digital converters (TDCs) and thus become power-hungry, while the aforementioned ASICs are not tested with a large-area SiPM array. These ASICs are not fully applicable to deep-water neutrino telescope experiments.

Thus, we are developing an ASIC to read out large-area SiPM arrays for TRIDENT. To identify a suitable readout circuit that satisfies the requirements, four different front-end circuits with optimized time resolution have been designed. Each architecture has its pros and cons. In addition, a four-channel digital summation mode is also verified. The most suitable architecture will be implemented in the fully functional ASIC. 

This article is organized as follows. The architectures are explained in section~\ref{sec:Architecture}, including the main modules. The layout of ASICs and test setups are presented in section~\ref{sec:setup}. Section~\ref{sec:results} provides performances of each circuit. Finally, the results are discussed and some conclusions are drawn in section~\ref{sec:conclusion}.

\section{Architecture}
\label{sec:Architecture}
\begin{figure}[htbp]
	\centering % \begin{center}/\end{center} takes some additional vertical space
	\includegraphics[width=.6\textwidth]{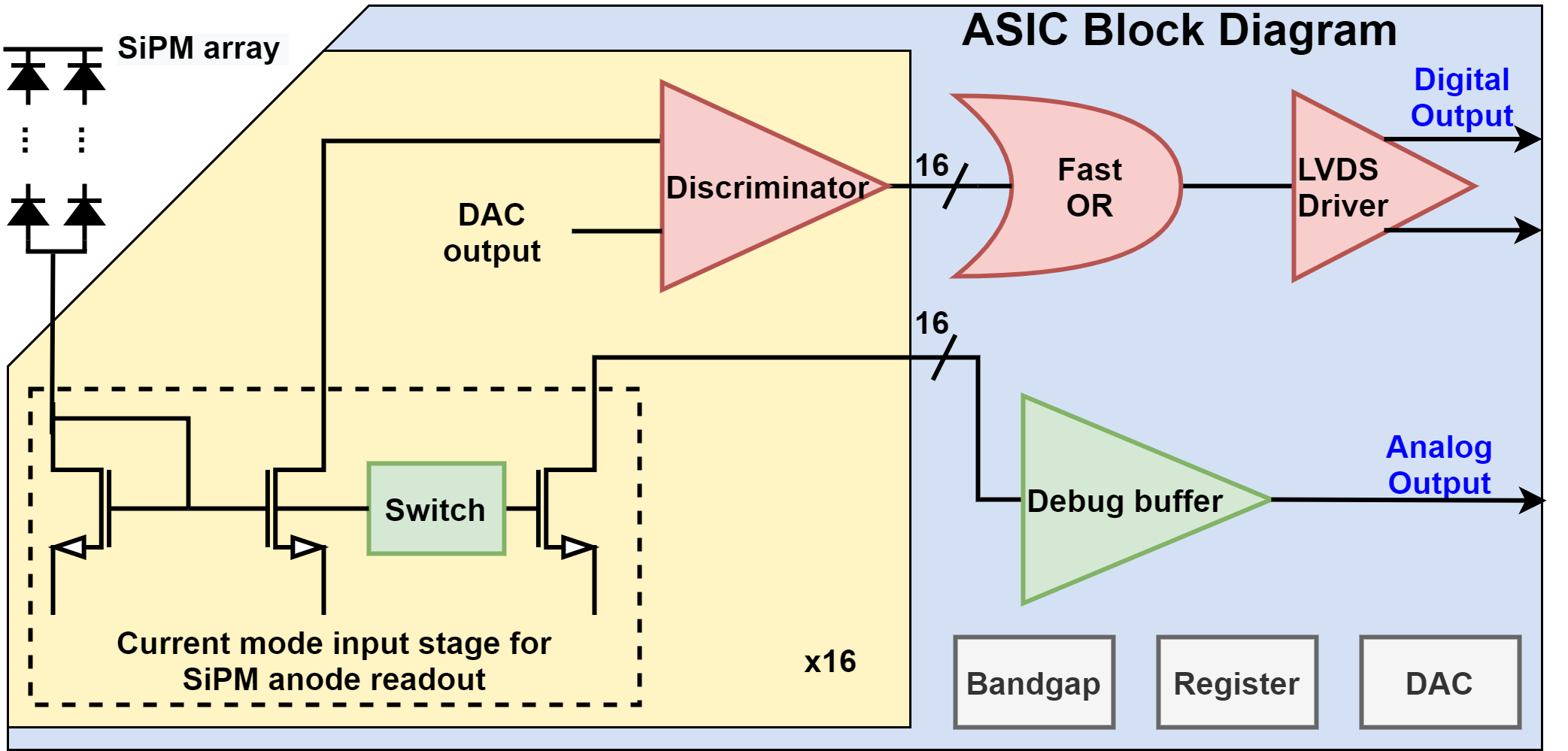}
	\caption{\label{fig:concept} Functional block diagram of the readout scheme.}
\end{figure}

The block diagram of the readout scheme is shown in figure~\ref{fig:concept}. Each input channel is connected to a SiPM array with two 3 $\times$ 3 mm$^{2}$ SiPMs in series and two series in parallel. The input stage can replicate the current signal of SiPMs as a current buffer and adjust the anode voltage to achieve bias voltage tuning. The input stage and discriminator are core modules, which are optimized for fast timing. The outputs of 16 discriminators are combined into one channel by a fast OR gate and subsequently output via a low-voltage differential signalling (LVDS) interface. The ASIC and SiPM array are abutted back to back on the base board and are distributed throughout the hDOM\cite{sipm1}. Such a summing method reduces the number of coaxial cables. Several auxiliary circuits have been designed, including analogue output buffers, digital-to-analogue converters (DACs) and a bandgap.

The specifications of the fully functional ASIC are summarized in table~\ref{tab:spec}. The time performance is optimized for Hamamatsu S13360-3050PE SiPMs \cite{s13360}. In order to achieve high-precision time measurement, different input stages and discriminators were designed and compared. The most suitable architecture will be used for the fully functional ASIC.

\begin{table}[htbp]
	\newcommand{\tabincell}[2]{\begin{tabular}{@{}#1@{}}#2\end{tabular}}
	\centering
	\caption{\label{tab:spec} The specification of the ASIC.}
	\smallskip
	\begin{tabular}{|l|l|}
		\hline
		\textbf{Parameter} & \textbf{Specification}\\
		\hline
		Input impedance & $\le$ 50 $\Omega$\\
		\hline
		Input connection & Anode connection\\
		\hline
		Anode voltage control per channel & About 1 V dynamic range adjustment\\
		\hline
		SPTR of a single channel & $\le$ 300 ps FWHM @ connecting a 3 $\times$ 3 mm$^{2}$ SiPM \\
		\hline
		Timing jitter & \tabincell{l}{$\le$ 500 ps FWHM @ connecting a 2s2p$^{*}$ array and \\ threshold is 0.5 photoelectron}\\
		\hline
		Output interface & LVDS\\
		\hline
		Power consumption & $\le$ 10 mW/channel\\
		\hline
		Number of channels & 16\\		
		\hline
	\end{tabular}
	\vspace{1ex}

	{\raggedright $^{*}$2s2p is a configuration with two 3 $\times$ 3 mm$^{2}$ SiPMs in series and two series in parallel.\par}
\end{table}

\subsection{Input stage}
\begin{figure}[htbp]
	\centering % \begin{center}/\end{center} takes some additional vertical space
	\includegraphics[height=.35\textwidth]{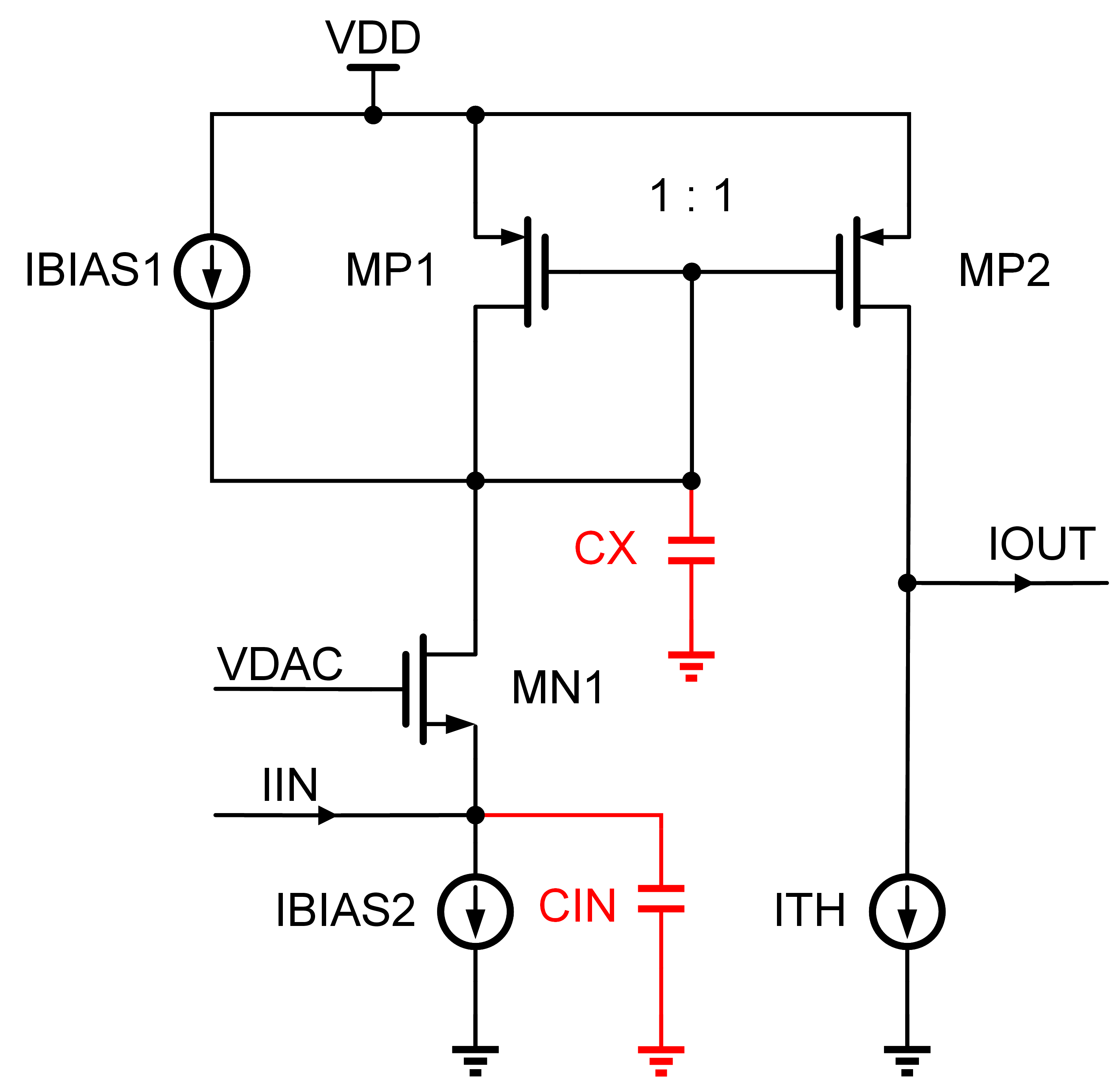}
	\qquad
	\includegraphics[height=.35\textwidth]{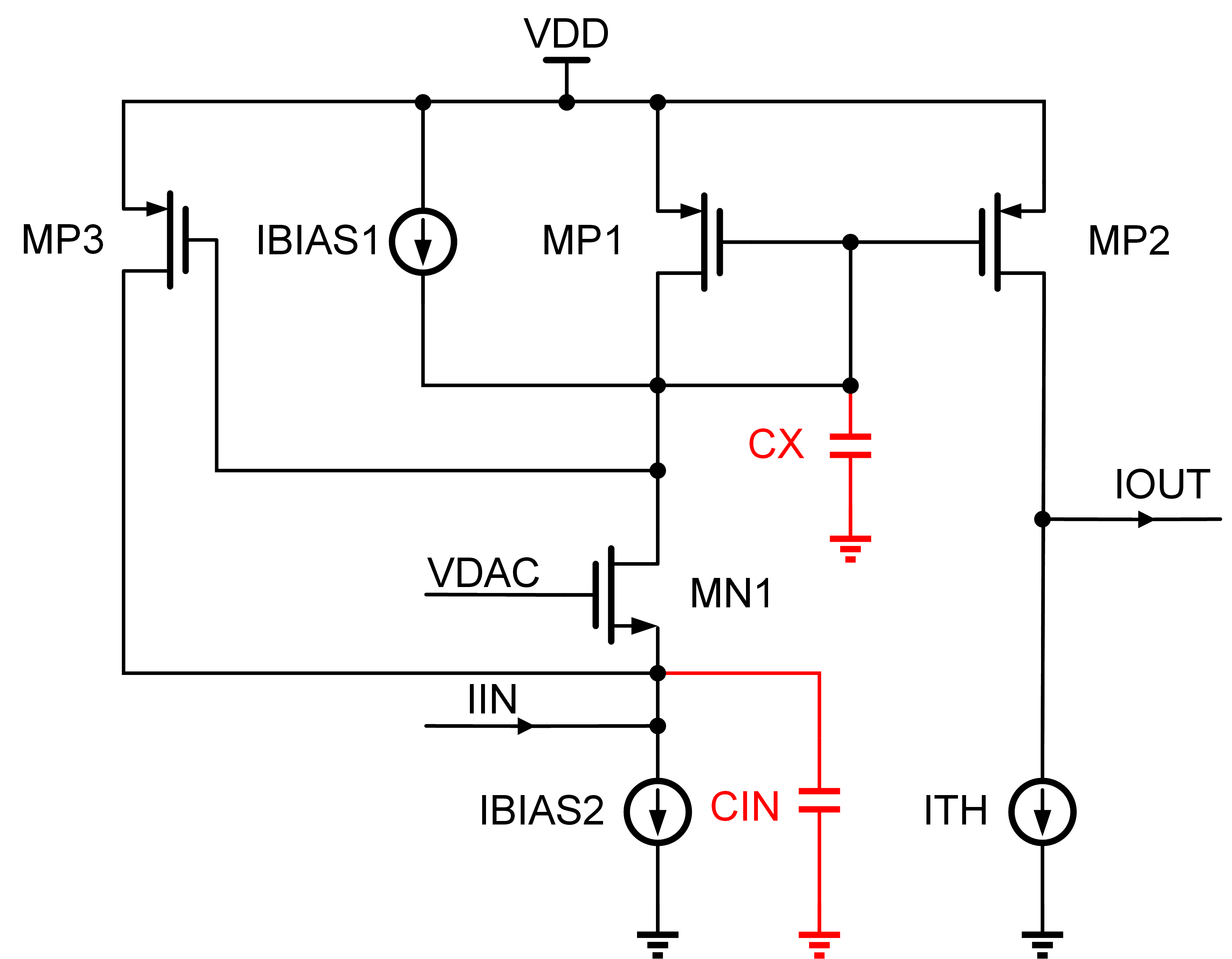}
	\caption{\label{fig:input} The schematics of two different current buffers. Left: common gate stage. Right: negative feedback common gate stage. CIN and CX represent the total capacitance of the node relative to the AC ground.}
\end{figure}

The input stage architectures include the common gate stage (CG) and the negative feedback common gate stage (NFBCG), shown in figure~\ref{fig:input}. The signal current is replicated by the current mirror and sent to the discriminator. Similar methods in \cite{DIET} are used for noise analysis, and the detail described in appendix section~\ref{sec:noise} shows that both architectures have similar time performance if the input capacitance is large. Compared to the CG, the NFBCG has a smaller input resistance. For this reason, the time constant of the falling edge and width of the output signal are smaller, which can alleviate the pile-up effect. Both architectures can adjust the anode voltage by changing the voltage of VDAC. The main characteristics of the input stages are shown in table~\ref{tab:inputspec}.
\begin{table}[htbp]
	\newcommand{\tabincell}[2]{\begin{tabular}{@{}#1@{}}#2\end{tabular}}
	\centering
	\caption{\label{tab:inputspec} Main characteristics of input stages.}
	\smallskip
	\begin{tabular}{|l|l|l|}
		\hline
		\textbf{Parameter} & \textbf{CG} & \textbf{NFBCG}\\
		\hline
		Input impedance & 50 $\Omega$ & 15 $\Omega$\\	
		\hline
		\tabincell{l}{Dynamic range of \\anode voltage adjustment}& About 1 V & About 1 V\\	
		\hline
		Bandwidth (without SiPMs) & Larger than 350 MHz & Larger than 500 MHz\\
		\hline
		Power consumption & About 3 mW & About 3 mW \\
		\hline
		Area & 2700 $\mu$m$^{2}$ & 4200 $\mu$m$^{2}$\\ 
		\hline
	\end{tabular}
\end{table}

\subsection{Discriminator}
The objective of the discriminator is to provide measurements of the ToA and time-over-threshold (ToT) of SiPM signals. The current signal output by the input stage can be directly timed by a current discriminator, or it can be converted into a voltage signal through a resistor and then timed by a voltage discriminator. Three discriminator architectures are designed. Two are current types and one is voltage type combined with a pre-amplifier.

\begin{figure}[htbp]
	\centering % \begin{center}/\end{center} takes some additional vertical space
	\includegraphics[height=.35\textwidth]{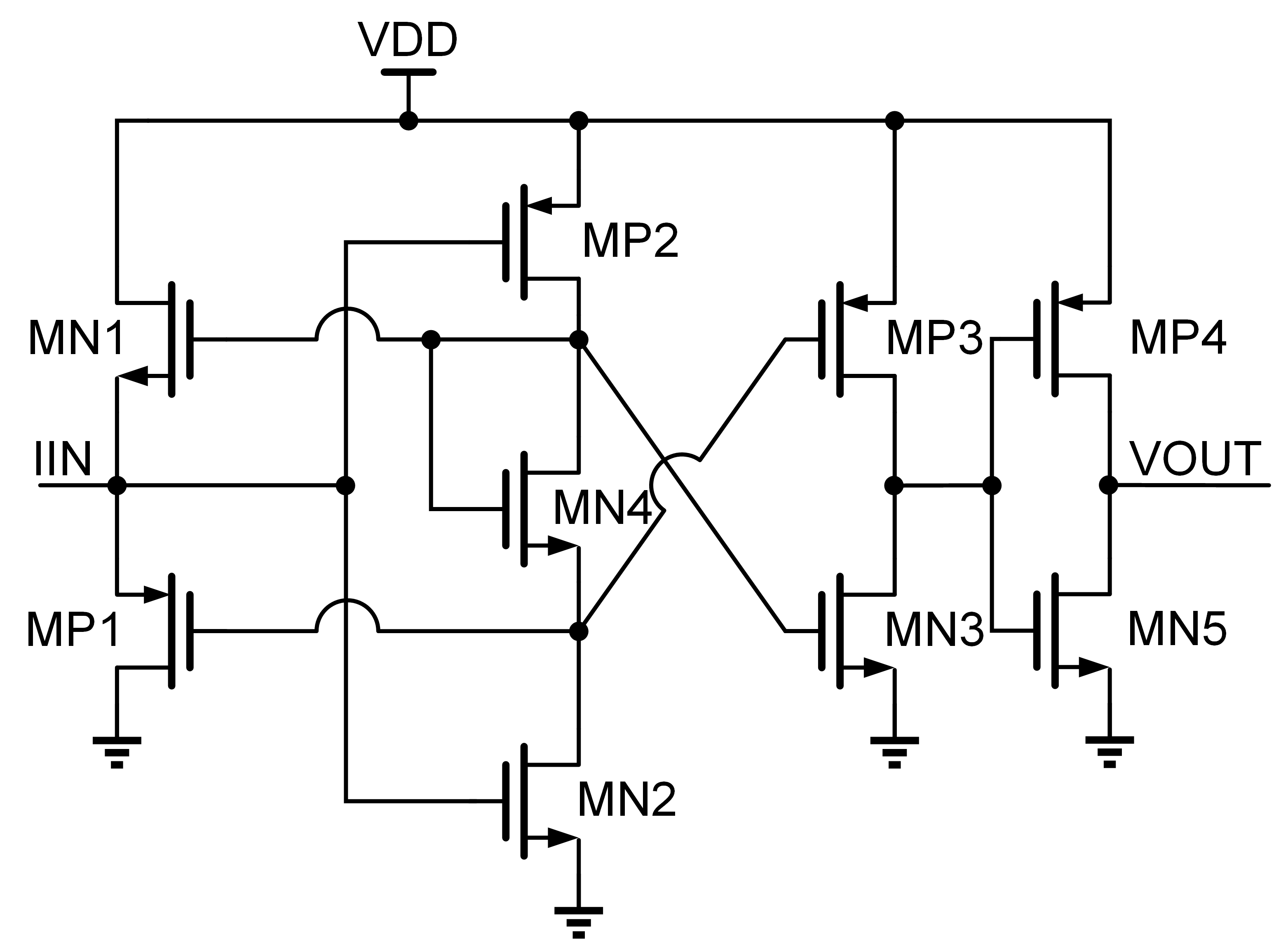}
	\qquad
	\includegraphics[height=.4\textwidth]{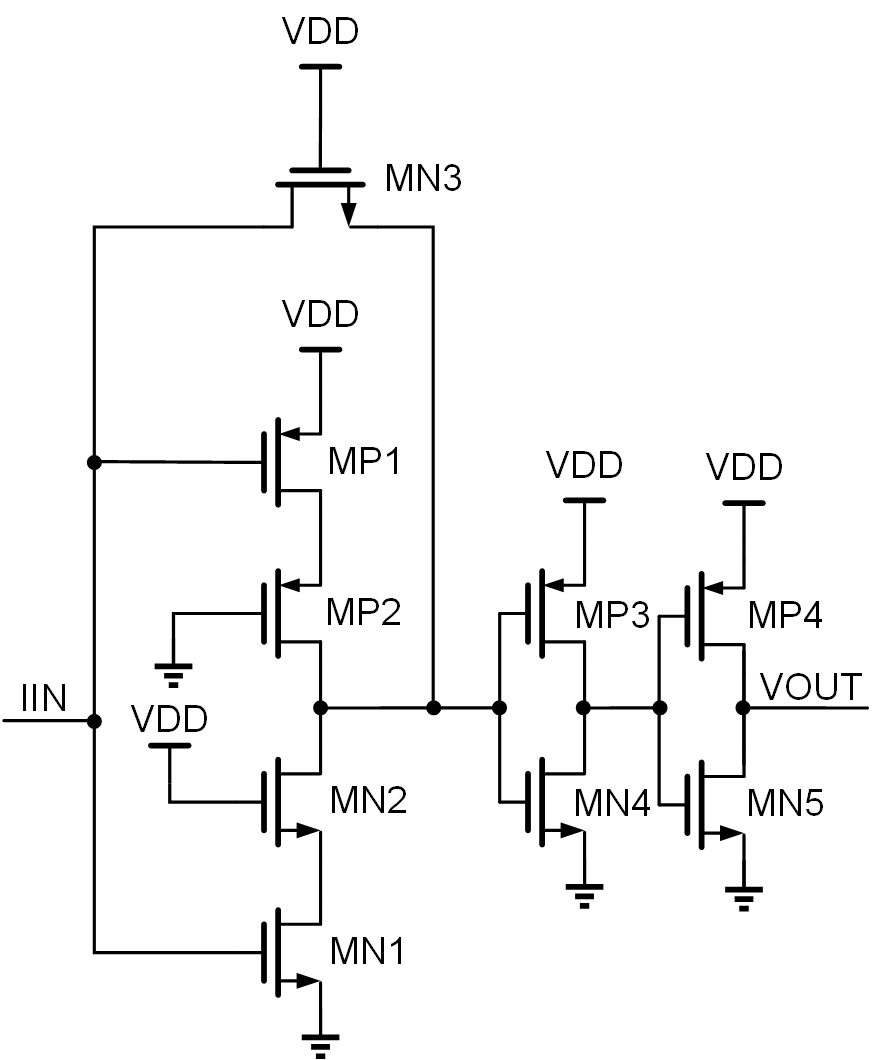}
	\caption{\label{fig:cdis1} The schematics of current discriminators. Left: current discriminators I. Right: current discriminators II.}
\end{figure}

In the first type of current discriminator \cite{disc1} (see the left panel of \figurename~\ref{fig:cdis1}), MP2 and MN2 form an inverter. The positive feedback formed by MP1 and MN1 reduces the input impedance and accelerates comparison. Diode-connected MN4 raises the gate voltage of MN1 to avoid the impact of MN1 and MN2 simultaneously turning off.
In the second type of current discriminator \cite{disc2} (see the right panel of \figurename~\ref{fig:cdis1}(b)), MP1 and MN1 form an inverter. MP2 and MN2 are used to reduce the bias current. MN3 works as a feedback resistor to reduce the input and output impedance, and thereby improve the response speed. 
The current injected into the current discriminator, IIN, is the difference between the output of the input stage and a current source. A DAC controls the current source to set the threshold.

\begin{figure}[htbp]
	\centering % \begin{center}/\end{center} takes some additional vertical space
	\includegraphics[height=.35\textwidth]{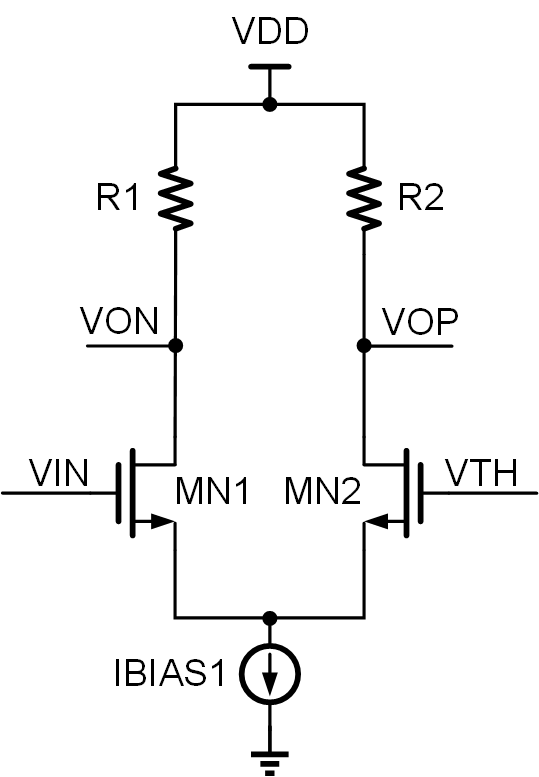}
	\qquad
	\includegraphics[height=.35\textwidth]{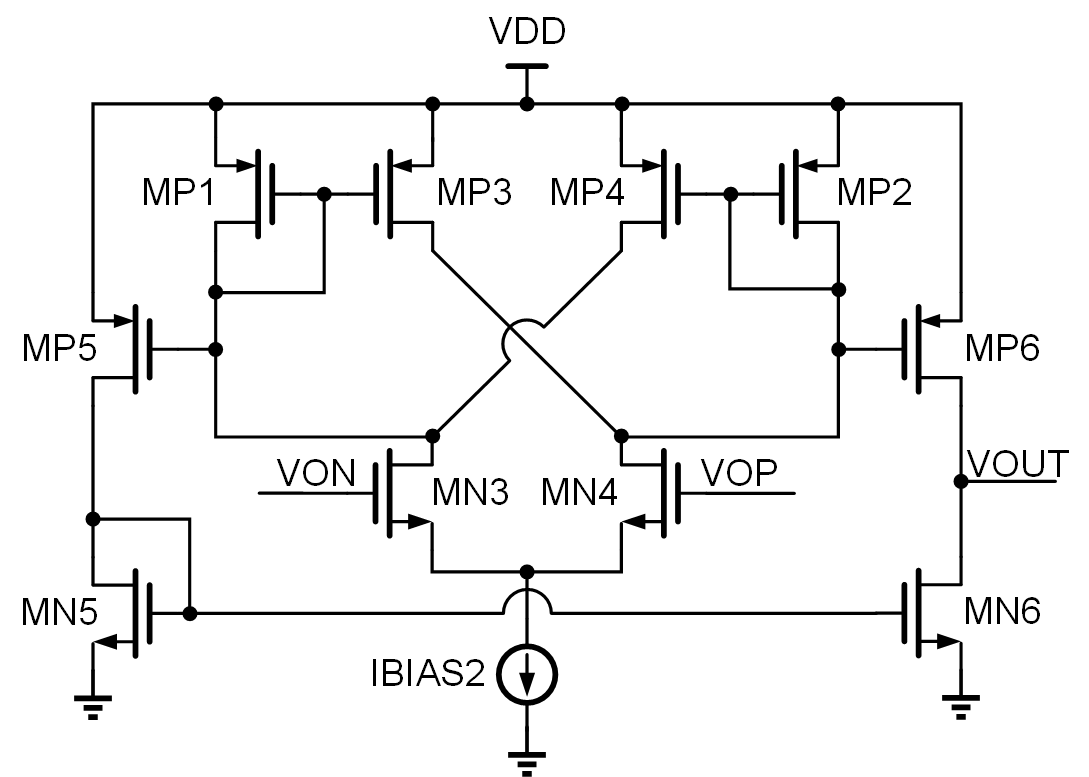}
	\caption{\label{fig:vdis} The schematics of the voltage discriminator. Left: one of the stages of the pre-amplifier. Right: hysteresis comparator.}
\end{figure}

As shown in \figurename~\ref{fig:vdis}, the voltage discriminator comprises a three-stage amplifier with resistive loads and a hysteresis comparator. The output of the current buffer is converted into a voltage signal by an on-chip resistor. The input stage is connected to the voltage discriminator via a series capacitor. 

The main characteristics of discriminators are listed in \tablename~\ref{tab:discspec}. The voltage discriminator has a higher area and power consumption. However, it is less affected by process-voltage-temperature (PVT) variations.
\begin{table}[htbp]\small
	\newcommand{\tabincell}[2]{\begin{tabular}{@{}#1@{}}#2\end{tabular}}
	\centering
	\caption{\label{tab:discspec} Main characteristics of discriminators.}
	\smallskip
	\begin{tabular}{|l|l|l|l|}
		\hline
		\textbf{Parameter} & \tabincell{c}{\textbf{Current discriminator} \\ \textbf{I}} & \tabincell{c}{\textbf{Current discriminator} \\ \textbf{II}} & \textbf{Voltage discriminator}\\
		\hline
		Input connection & DC coupling & DC coupling & AC coupling\\
		\hline
		Propagation delay & 0.56 ns	& 0.47 ns & 1.9 ns\\
		\hline
		\tabincell{l}{Static power \\consumption} & 0.6 mW & 0.4 mW & 3.1 mW\\
		\hline
		Area & 187 $\mu$m$^{2}$	& 336 $\mu$m$^{2}$ & 6000 $\mu$m$^{2}$\\
		\hline
	\end{tabular}
\end{table}

\subsection{Digital summation}
In order to reduce the number of output interfaces, it is important to perform a summation of the signal read out from different SiPMs in an array. As mentioned in \cite{summation}, two strategies of combining the signals of each input stage are applicable. The first option involves an analogue summation of all the output signals collected by the input stage prior to the discriminator, which is applied in MUSIC ASIC \cite{MUSIC}. Then, a TDC digitizes the output of the discriminator. Analogue circuits are generally large in area, and the added parasitic parameters will deteriorate the time performance. The second option equips each channel with an on-chip TDC and then uses a digital module to merge the data of each TDC. For high-precision time measurement, a large number of TDCs will significantly increase power consumption and area. We chose a digital summation similar to that in \cite{EXYT} and replaced the wire-OR module with a fast CMOS OR gate. Hence, the rising and falling speeds of the signal are balanced. The summed signal drives an LVDS driver to output. The time measurements are finally completed by an FPGA-based TDC.

\section{ASIC layout and test setups}
\label{sec:setup}
\begin{figure}[htbp]
	\centering % \begin{center}/\end{center} takes some additional vertical space
	\includegraphics[width=0.8\textwidth]{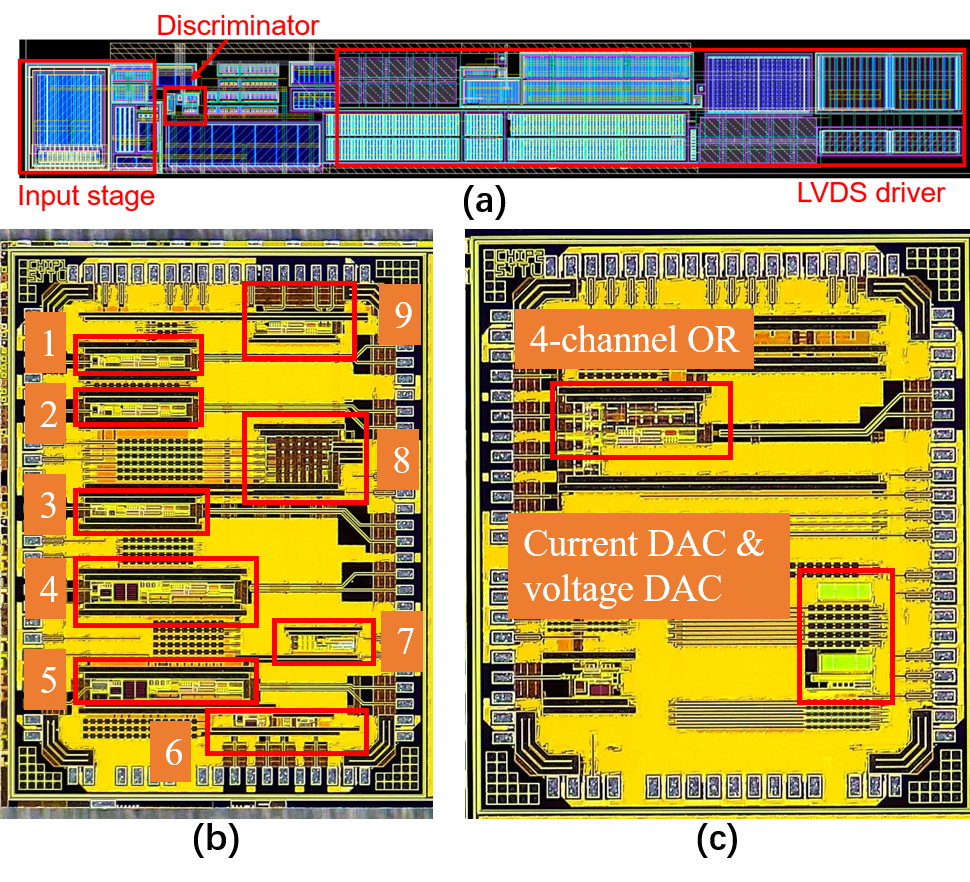}
	\caption{\label{fig:layout} (a) Single timing channel layout. (b) The microscope photo of chip 1, including T11 (1), T12 (2), T22 (3), T1V (4), ESD-enhanced T1V (5), input stages with analogue output buffers (6), bandgap (7), bias (8), and stand-alone LVDS driver (9). (c) The microscope photo of chip 2.}
\end{figure}

Two ASICs have been designed for testing different circuits and fabricated using a 180~nm CMOS process. Four timing channels are named T11 (combination of CG and current discriminator I), T12 (combination of CG and current discriminator II), T22 (combination of NFBCG and current discriminator II), and T1V (combination of CG and voltage discriminator). An OR gate sums the outputs of four T12s. Additionally, analogue front-end circuits are placed in deep N wells for noise isolations. A timing channel layout and microscope photos of chips are present in \figurename~\ref{fig:layout}. Chip 1 has 84 pins with a die area of $1.92 \times 2.56$  mm$^{2}$ and is packaged in LQFP100 (low profile quad flat package). Chip 2 has 72 pins with a die area of $1.84 \times 2.16$  mm$^{2}$ and is packaged in LQFP80.

\begin{figure}[htbp]
	\centering % \begin{center}/\end{center} takes some additional vertical space
	\includegraphics[width=0.8\textwidth]{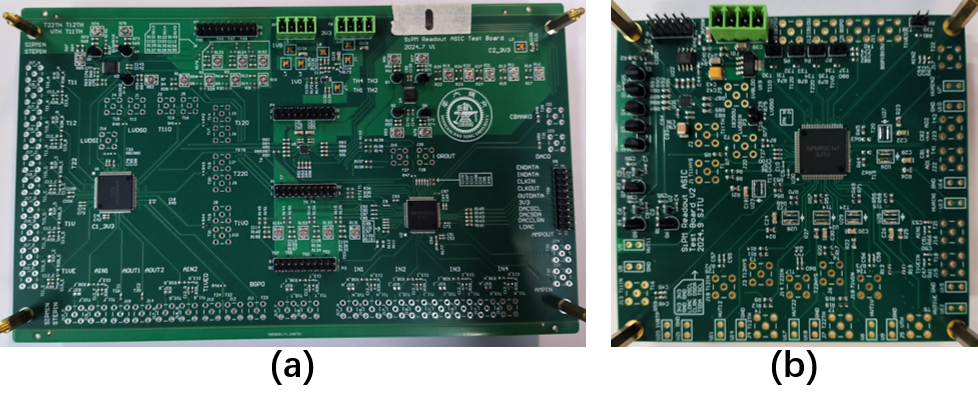}
	\caption{\label{fig:pcb} (a) The first version of the evaluation board contains mainly chip 1 and chip 2. The board measures $228\times127$ mm$^{2}$. (b) The second version of the evaluation board contains mainly chip 1. The board measures $107\times99$ mm$^{2}$.}
\end{figure}

\begin{figure}[htbp]
	\centering % \begin{center}/\end{center} takes some additional vertical space
	\includegraphics[width=0.6\textwidth]{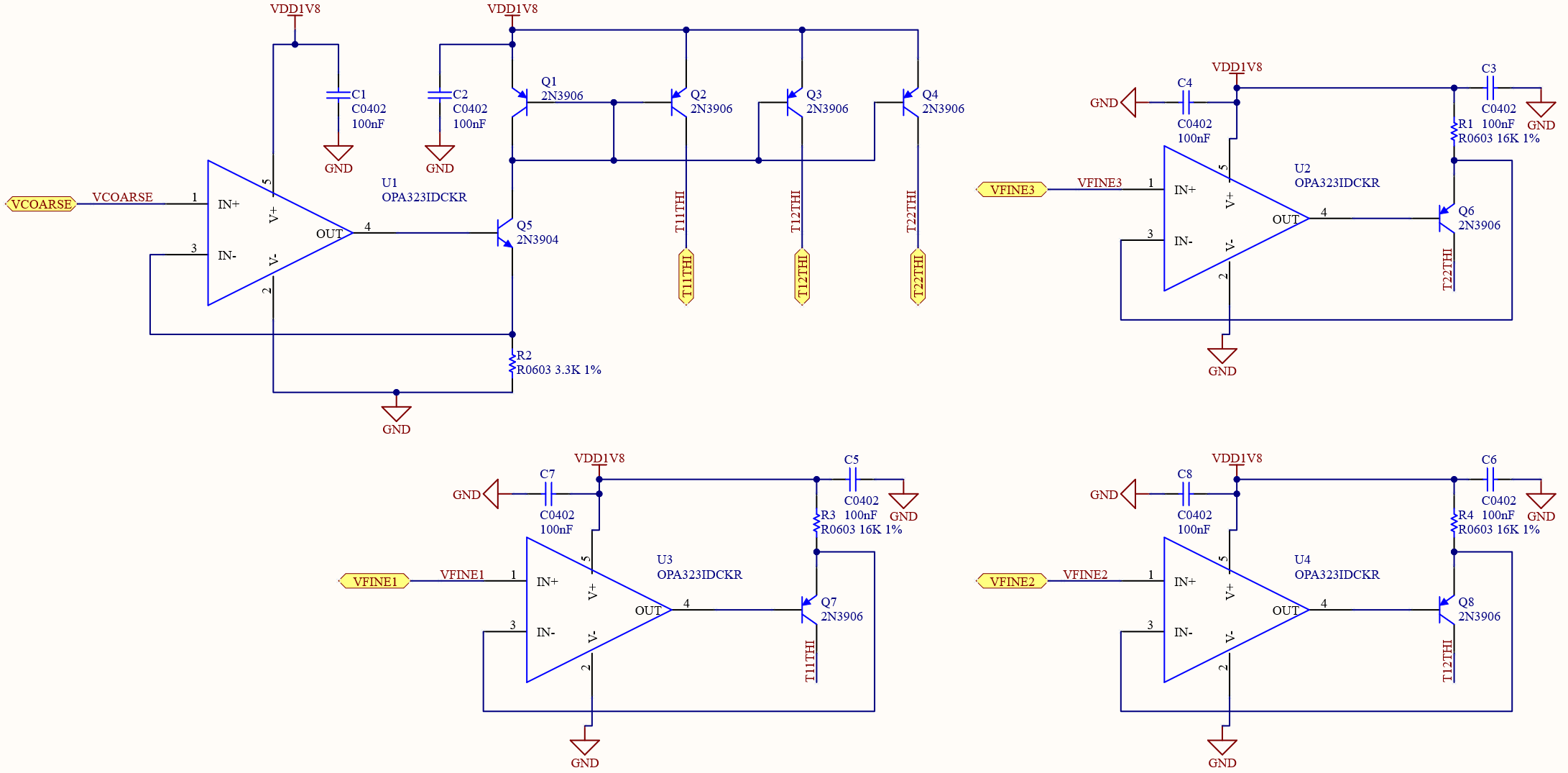}
	\quad
	\includegraphics[width=0.3\textwidth]{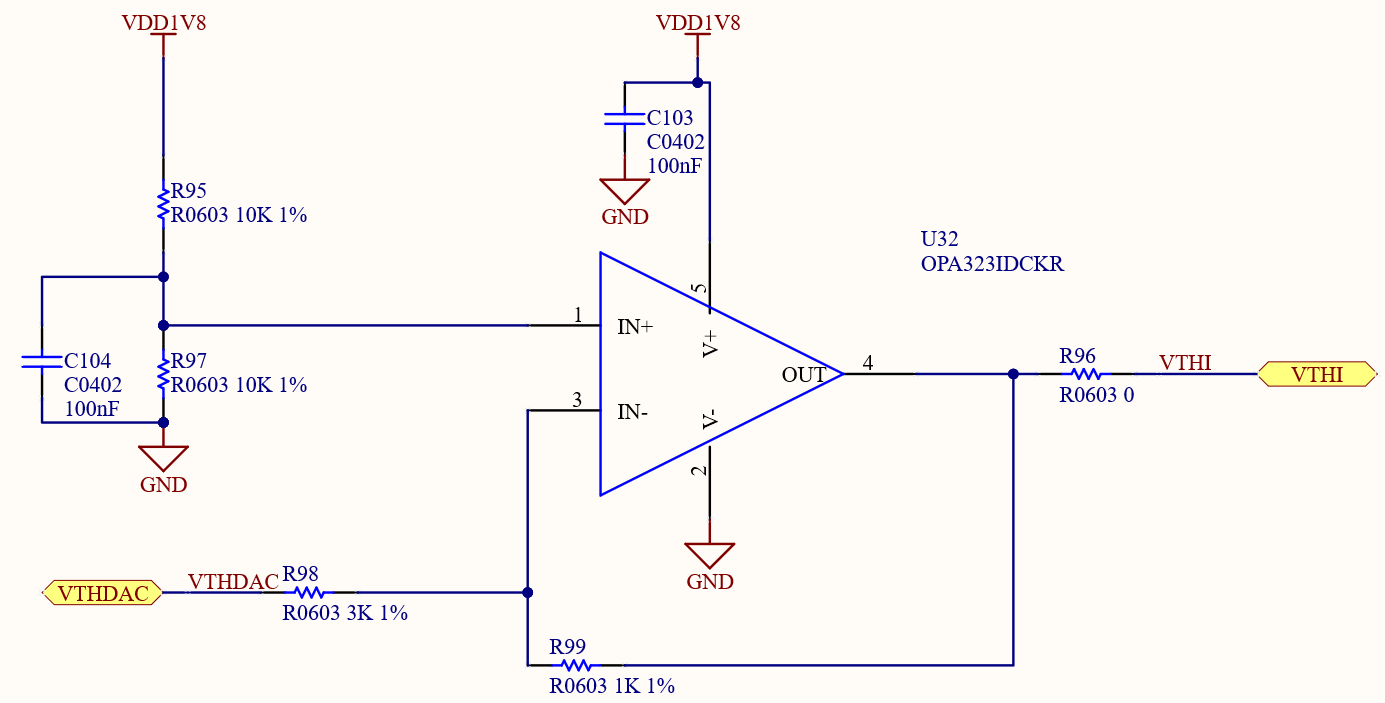}
	\caption{\label{fig:thcir} Left: the current threshold set circuit. VCOARSE, VFINE1, VFINE2, and VFINE3 are outputs of the DAC. The thresholds of T11, T12, and T22 are set by T11THI, T12THI, and T22THI respectively. Right: the voltage threshold set circuit. VTHDAC is an output of the DAC. The thresholds of T1V and ESD-enhanced T1V are set by VTHI.}
\end{figure}

Two versions of the evaluation board were developed, as shown in \figurename~\ref{fig:pcb}. The main difference between the two versions is the set of the discriminator threshold. For the first version, the threshold could be set by off-chip variable resistors (Bourns Inc. TC33) or a power source (Keysight B2962B). The former is inadequate for threshold scan testing and the latter requires additional complicated programming. 
The circuits for the threshold set of the second version board are illustrated in \figurename~\ref{fig:thcir}. An 8-channel DAC (Texas Instruments DAC53608) provides adjustable reference voltages. The current thresholds are set by a voltage-to-current circuit with a two-stage adjustment. The coarse adjustment range is 100 $\mu$A to 300 $\mu$A with a step of 2 $\mu$A. The fine adjustment range is 0 $\mu$A to 50 $\mu$A with a step of 0.44 $\mu$A. The reference voltage is used as the threshold voltage after linear transformation by the operational amplifier circuit (Texas Instruments OPA323) with an adjustment range of 0.6 V to 1.2 V and a step of 0.59 mV.

\begin{figure}[htbp]
	\centering % \begin{center}/\end{center} takes some additional vertical space
	\includegraphics[width=0.6\textwidth]{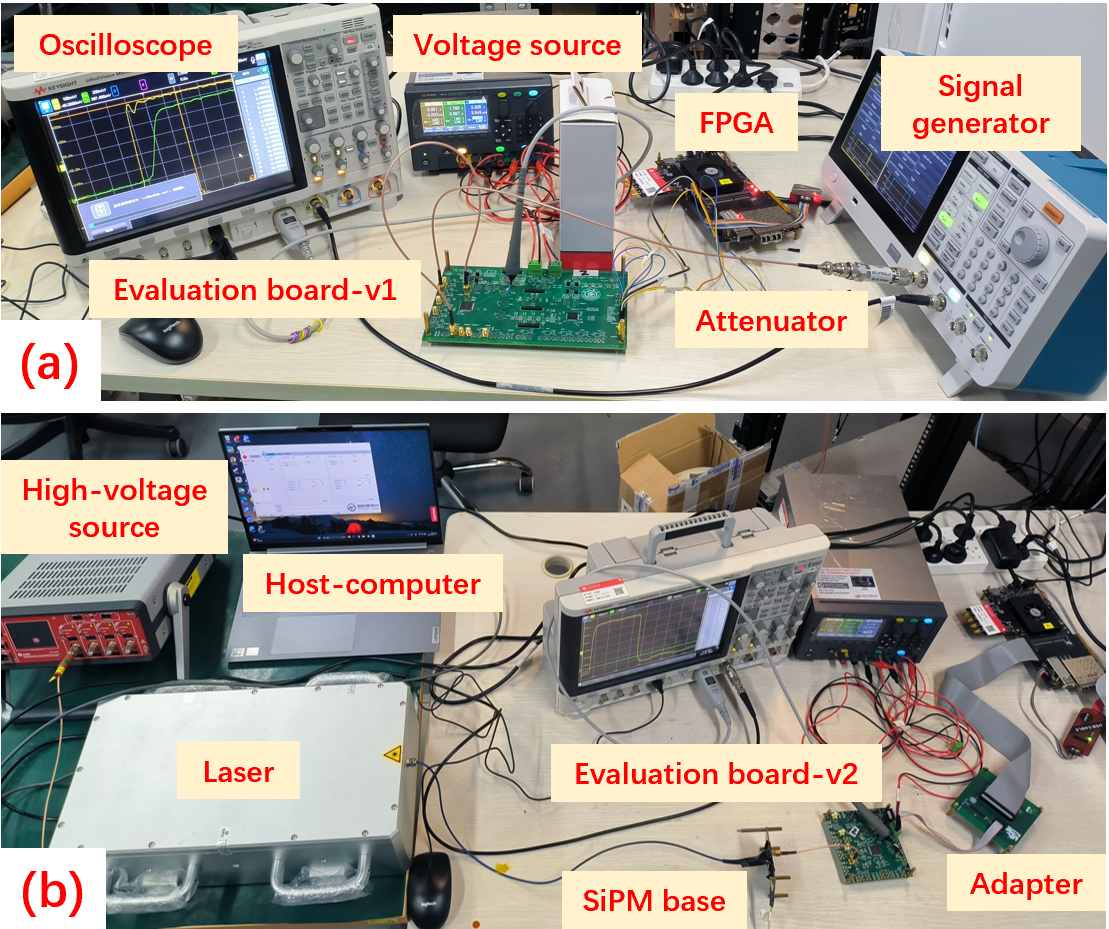}
	\caption{\label{fig:setup1} (a) System for the electronic test. (b) System for the SiPM commissioning. SiPMs are mounted on the base, and the light-tight box for SiPMs is removed.}
\end{figure}

To entirely evaluate the time performance of ASICs, two test systems have been set up. As shown in \figurename~\ref{fig:setup1} (a), the first system measures the electronics performance of ASICs only. The DC power is supplied by a voltage source (Keysight E36312A). The off-chip DAC is configured by an FPGA board (ALINX AX7325) to set bias voltages. 
A differentiating circuit \cite{DIET} converts a step voltage signal to an exponential decay current signal. The injected voltage signal is from a signal generator (Tektronix AFG31252) and attenuated by a 30-dB attenuator (Keysight 8491B). The output of the LVDS driver is acquired by a Keysight MSOX6004A 2.5 GHz oscilloscope (20 GS/s) with a Keysight N2750A differential probe (1.5 GHz). Meanwhile, another synchronized channel of the signal generator is connected to the oscilloscope as a trigger.

\figurename~\ref{fig:setup1} (b) presents the setup for the SiPM commissioning which is mainly used to measure the SPTR of the whole readout. The SiPM is mounted on a base and its anode is connected to the evaluation board via a 10 cm coaxial cable. The bias voltage of the SiPM is provided by a high-voltage source (CAEN DT1470ET). A femtosecond laser (PulseX Laser Technology Explorer-FGL-5, central wavelength at 532 nm, pulse duration < 1 ps, and jitter < 1 ps) emits a pulse of light which is attenuated to 0.01\% before reaching the SiPM. The intensity of the laser is controlled by the host-computer so that most outputs of SiPM are signal photoelectron (SPE) signals. The trigger of the oscilloscope is generated synchronously by the laser.

\section{Test results}
\label{sec:results}
The goal of this section is to illustrate the time performance of the ASIC through various experimental tests. A total of seven chips 1 and five chips 2 have been randomly selected and tested, and their performances are similar. 

\begin{figure}[htbp]
	\centering % \begin{center}/\end{center} takes some additional vertical space
	\includegraphics[width=0.4\textwidth]{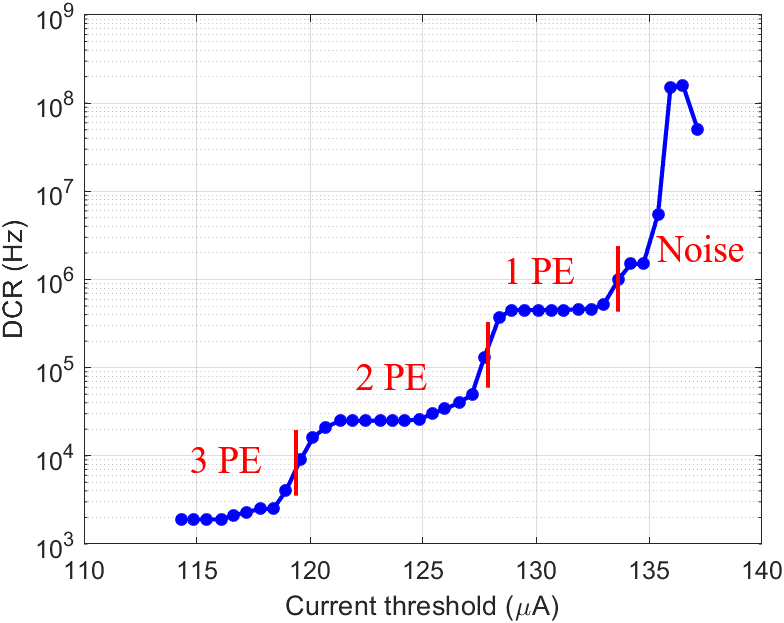}
	\qquad
	\includegraphics[width=0.4\textwidth]{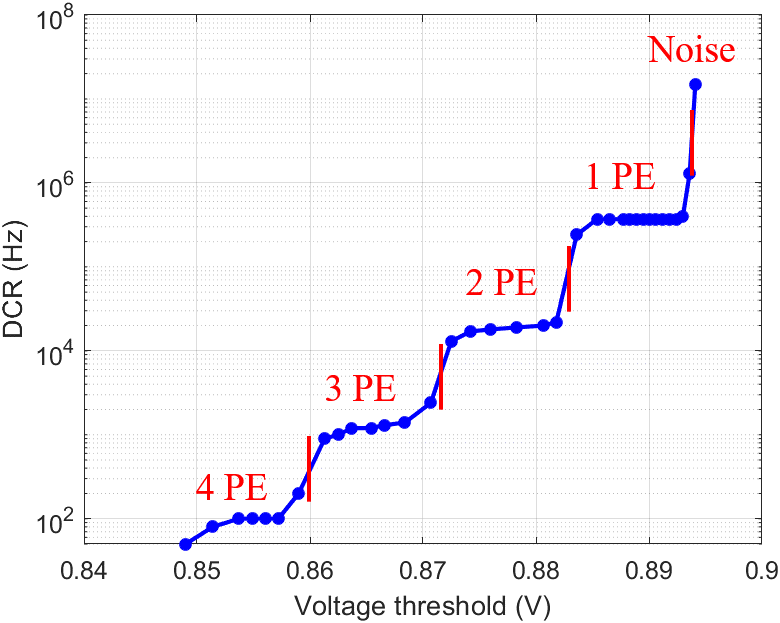}
	\caption{\label{fig:darkrate} Measured dark rate at different thresholds. Each photoelectron amplitude's location can be identified at the edges where the rate sharply declines. Left: an example of timing channels with the current discriminator. Right: an example of timing channels with the voltage discriminator.}
\end{figure}

Due to the process variation, the thresholds of different timing channels have a range of deviations. The calibration is performed by recording the dark count rate (DCR) of the SiPM when sweeping the threshold \cite{DIET2}. The typical curve of DCR changing with the threshold value is shown in \figurename~\ref{fig:darkrate}. The 0.5 photoelectron (PE) threshold can be set at the middle point of the SPE step. Based on the simulations and measurements, the required DAC bit numbers for T11, T12, T22, and T1V are 11, 11, 6, and 5, respectively.

\subsection{Test of input stages}
\begin{figure}[htbp]
	\centering % \begin{center}/\end{center} takes some additional vertical space
	\includegraphics[width=0.4\textwidth]{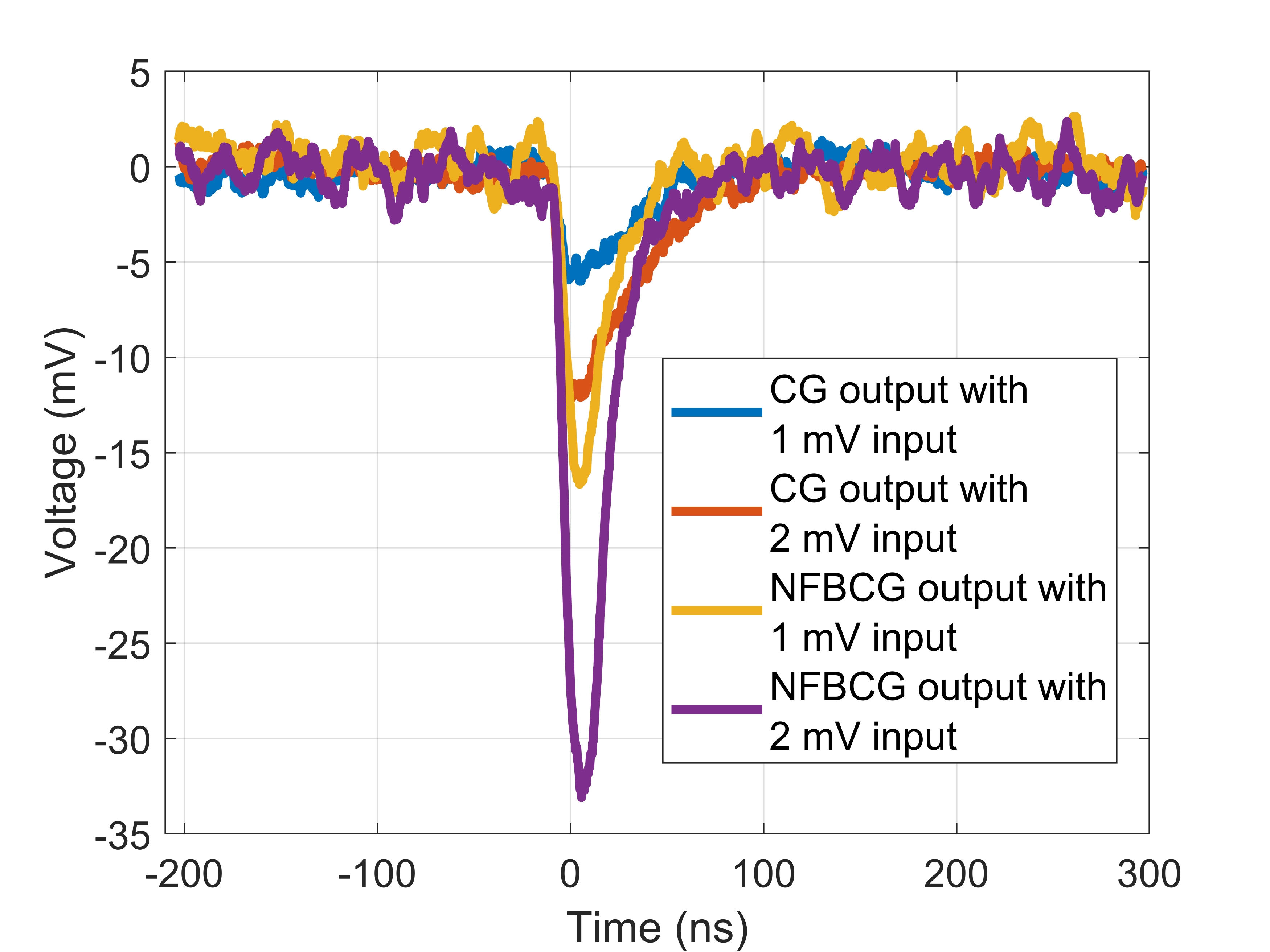}
	\qquad
	\includegraphics[width=0.4\textwidth]{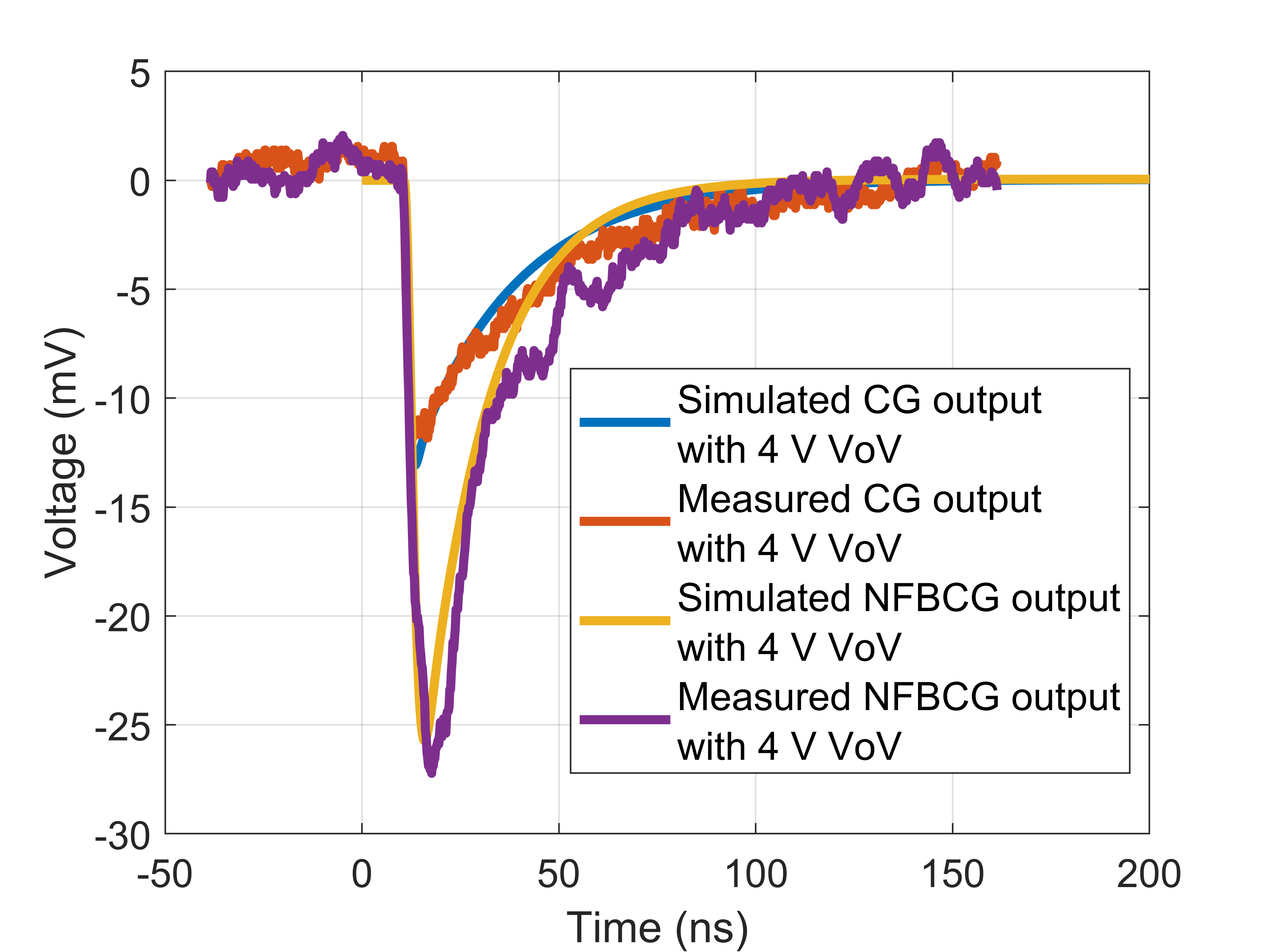}
	\caption{\label{fig:debugout} Left: waveforms of input stages with the signal generator input, where the input amplitude is after attenuation. Right: simulated and measured SPE waveforms of the input stage readout, where the VoV of the SiPM is 4~V.}
\end{figure}

As shown in \figurename~\ref{fig:layout} (b), input stages can be tested independently. The waveforms are shown in \figurename~\ref{fig:debugout}, including injections of the signal generator and inputs of the SiPM SPE. In addition, the simulated results with the SiPM electrical model \cite{sipmmodel} are present, which are consistent with the measurements. Considering the trade-off between the gain and the dark rate, SiPMs under simulation and test were typically biased at an over voltage (VoV) of 4 V.
As can be observed, the outputs of the CG and the NFBCG have the same slopes. However, NFBCG outputs have larger amplitudes and shorter decay times, which is in agreement with the calculation in section~\ref{sec:noise}.

The current signals output from input stages are converted into voltage signals without amplifications via an on-chip resistor. Consequently, the amplitude of original SPE signals can be deduced, which is consistent with the result obtained by the threshold scanning. 

The mapping between the signal generator injection signal and the SiPM PE signal is obtained by analysing the simulation and measurement results of input stages. For instance, the slope of the SPE signal output by the input stage is consistent with the slope of the 6 mV (value after attenuation) signal injected by the signal generator.

\subsection{Test of the electronic jitter of ASICs}
\label{sec:jitter}

\begin{figure}[htbp]
	\centering % \begin{center}/\end{center} takes some additional vertical space
	\includegraphics[width=0.4\textwidth]{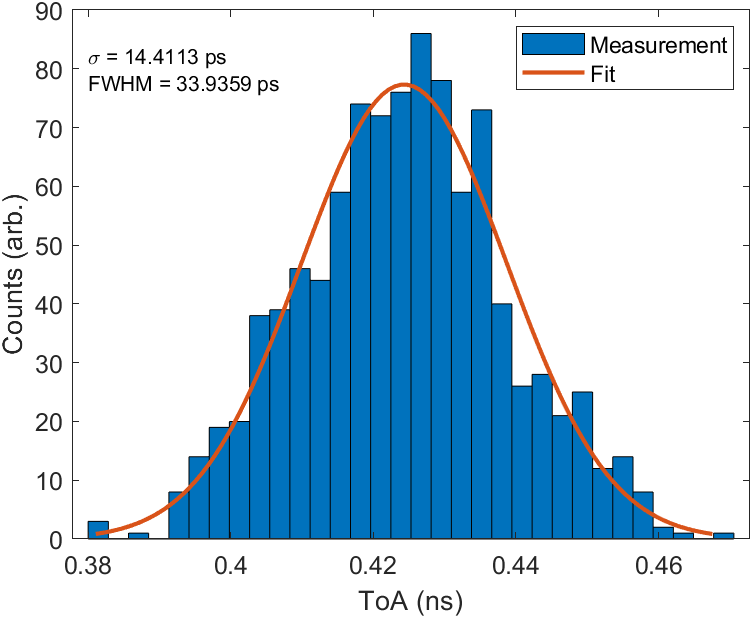}
	\qquad
	\includegraphics[width=0.4\textwidth]{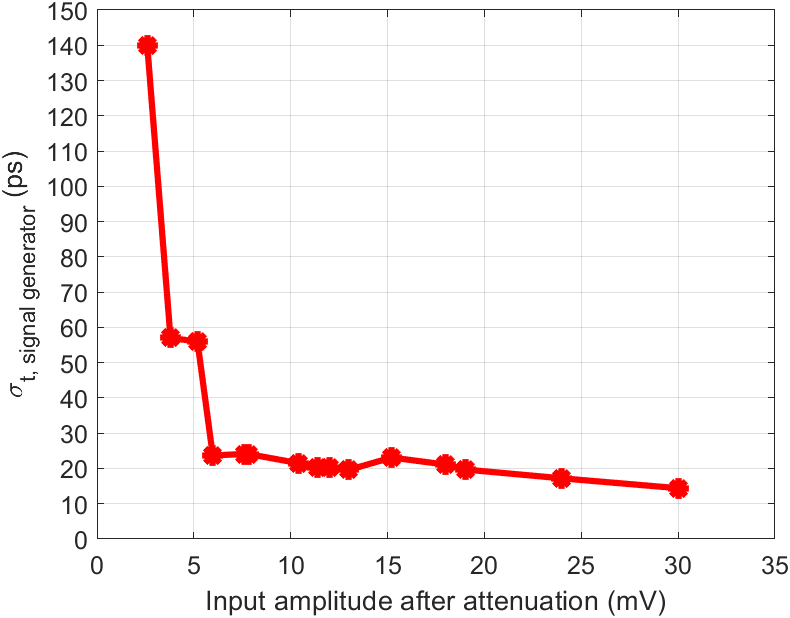}
	\caption{\label{fig:sigjitter} Test results of the jitter of the signal generator. Left: an example of the ToA distribution with a Gaussian fit. Right: the curve of jitter changing with the output amplitude.}
\end{figure}

The time difference between the output and the trigger is taken as ToA. First of all, the jitter of the signal generator is measured as shown in \figurename~\ref{fig:sigjitter}. The jitter of ASICs is subtracted using equation \eqref{eq:41}.
\begin{equation}
	\label{eq:41}
	\begin{split}
		\sigma_{t,ASIC}=\sqrt{\sigma_{t,ToA}^{2}-\sigma_{t,signal\,generator}^{2}}
	\end{split}
\end{equation}
where $\sigma_{t,ToA}$ is the standard deviation of the measured ToA and $\sigma_{t,signal\,generator}$ is the jitter of the signal generator.

\begin{figure}[htbp]
	\centering % \begin{center}/\end{center} takes some additional vertical space
	\includegraphics[width=0.4\textwidth]{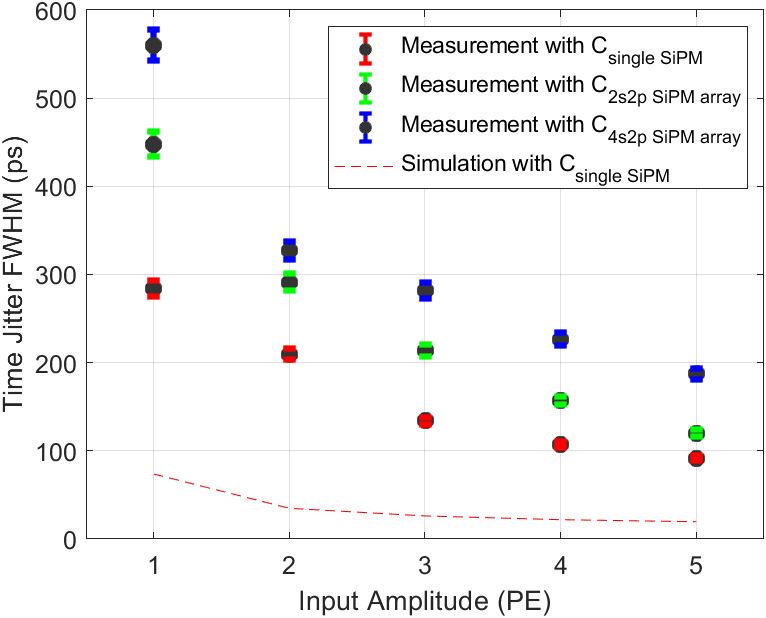}
	\qquad
	\includegraphics[width=0.4\textwidth]{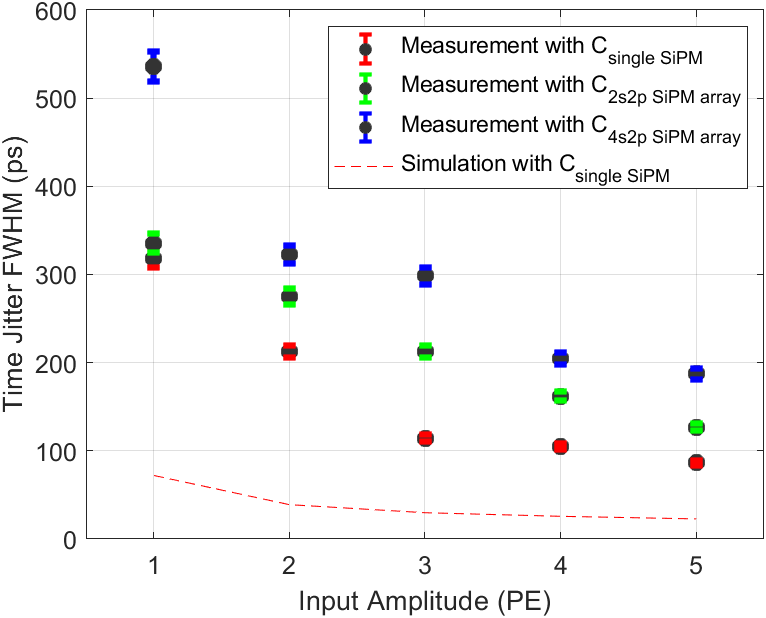}
	\qquad
	\includegraphics[width=0.4\textwidth]{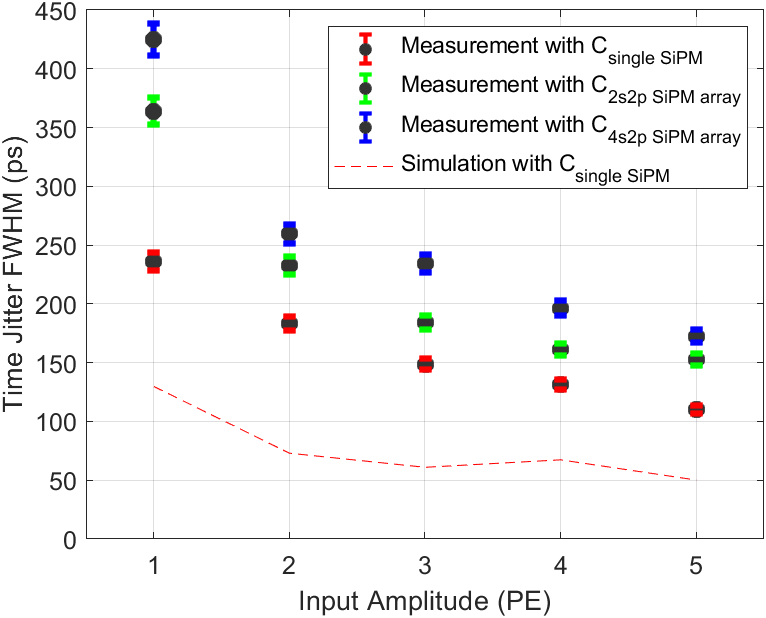}
	\qquad
	\includegraphics[width=0.4\textwidth]{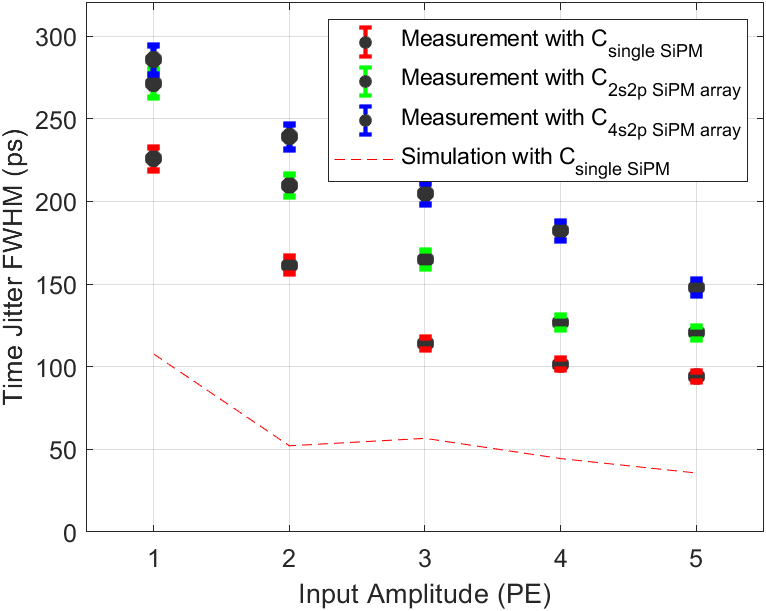}
	\caption{\label{fig:ejitter} Timing channel jitter versus input amplitude: T11 (top left), T12 (top right), T22 (bottom left), and T1V (bottom right).}
\end{figure}

The aforementioned mapping method and differentiating circuits are applied to obtain the electronic jitter of the ASICs with different SiPM arrays connected to the input. The test results of the time jitter with ASICs only are shown in \figurename~\ref{fig:ejitter} attaching simulated results, where ``nsmp'' indicates a configuration with ``n'' 3$\times$3 mm$^{2}$ SiPMs in series and ``m'' series in parallel.
Several noise sources and parasitic parameters, including power supply noise and parasitics of the printed circuit board, were ignored during the simulation. For this reason, the simulated results are much better. The time jitter of four architectures fulfils the specifications. In particular, the timing performance of T1V is significantly superior when connecting large area arrays.

Each channel of the 4-channel OR circuit has been tested with $C_{single\,SiPM}$. For instance, the mean values of the ToA distributions of the four channels are 3.543~ns, 3.341~ns, 3.268~ns, and 3.469~ns, and the standard deviations are 109~ps, 95 ~ps, 97~ps, and 107~ps, respectively. As a result, the time performance of stand-alone channels in a 4-channel OR circuit is similar to separate timing channels. The total electronic jitter of the four channels increases slightly after the digital summation, with a value of 350 ps FWHM.

\subsection{SPTR measurements}
\begin{figure}[htbp]
	\centering % \begin{center}/\end{center} takes some additional vertical space
	\includegraphics[width=0.4\textwidth]{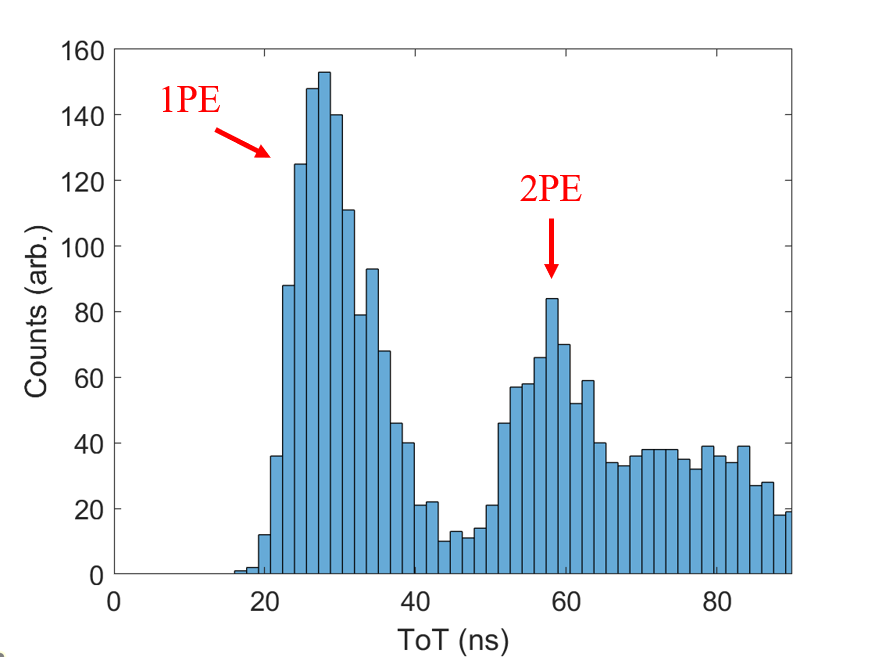}
	\qquad
	\includegraphics[width=0.4\textwidth]{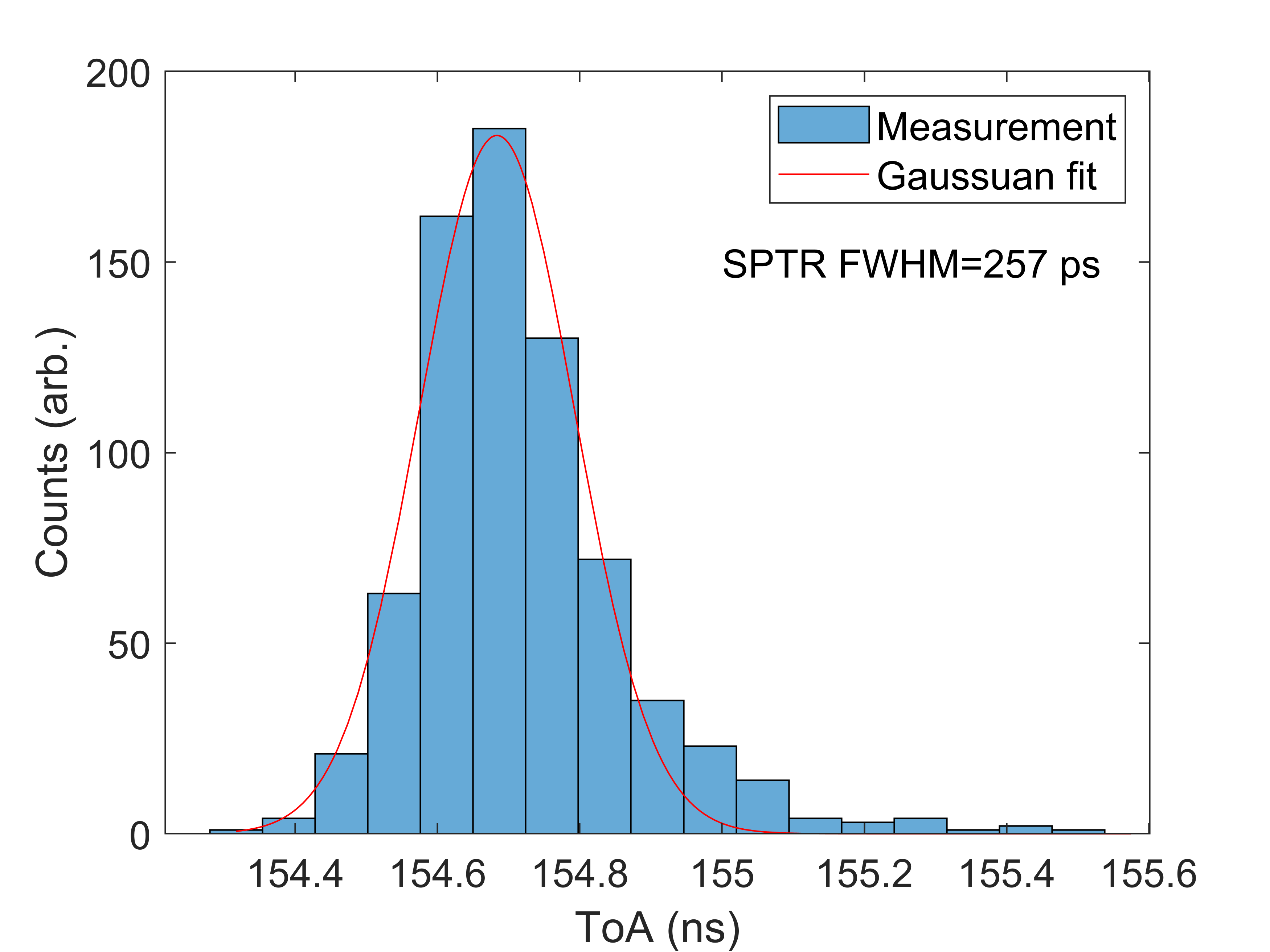}
	\caption{\label{fig:sptr1} Left: an example of the ToT distribution. The peak corresponding to SPEs can be clearly seen. Right: an example of the ToA of PE signals distribution with a Gaussian fit.}
\end{figure}

\begin{figure}[htbp]
	\centering % \begin{center}/\end{center} takes some additional vertical space
	\includegraphics[width=0.5\textwidth]{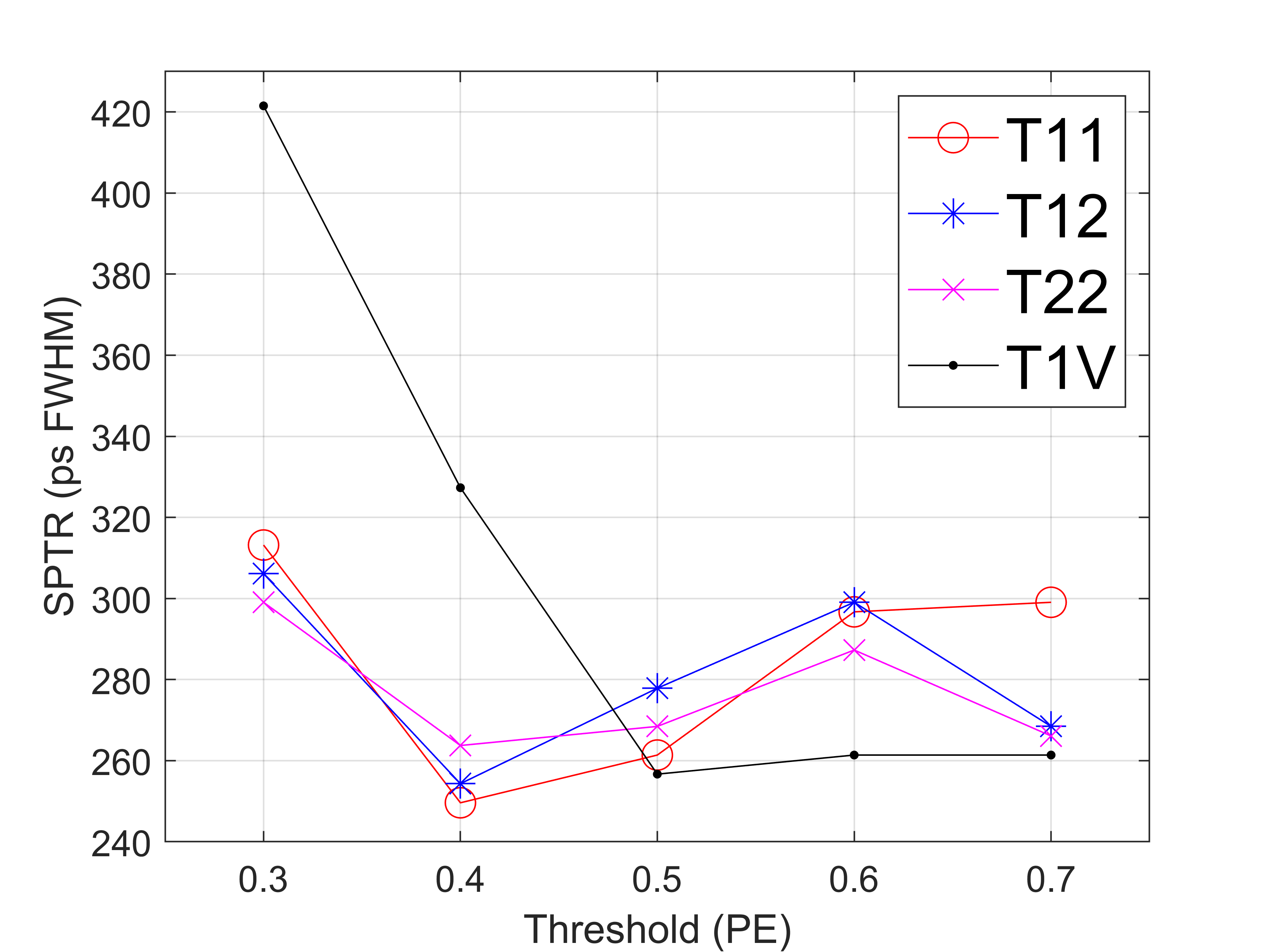}
	\caption{\label{fig:sptr2} SPTR results for different architectures with different thresholds.}
\end{figure}

The results of SPTR measurements can be seen in \figurename~\ref{fig:sptr1} and \figurename~\ref{fig:sptr2} for the Hamamatsu S13360-3050PE SiPM with 4 V of VoV.
The SPTR performance of all four architectures is qualified and similar at the 0.5 PE threshold.

\section{Discussion and conclusion}
\label{sec:conclusion}
This work researches the performance of four architectures of large-area SiPM array readout circuits, which present lower power consumption (less than 7 mW/channel) and satisfying time performance. All of them meet requirements that 3~$\times$~3~mm$^{2}$ SiPM SPTR is less than 300~ps FWHM and 2s2p SiPM array (two 3 $\times$ 3 mm$^{2}$ SiPMs in series and two series in parallel) timing jitter is less than 500~ps FWHM. 
When the threshold is set to 0.5~PE and the input is connected to a 2s2p SiPM array, the timing performance of CG is slightly better than that of NFBCG. Since the impact of PVT variation on the voltage discriminator is relatively minor, it is more appropriate than the current discriminator. 

A comparison of time performances with other SiPM readout ASICs is summarized in \tablename~\ref{tab:performance}. NINO has excellent time performance, while it consumes more power. Petiroc2A was tested under higher bias voltage for the SiPM, at the expense of increased DCR. The performance of our four different architectures is similar to that of HRFlexToT, FlexToT and TOFPET2. The experimental measurements have demonstrated the potential of designed circuits for fast-timing applications with large-area SiPM arrays, and they are the baseline for future design.
\begin{table}[htbp]
	\newcommand{\tabincell}[2]{\begin{tabular}{@{}#1@{}}#2\end{tabular}}
	\centering
	\caption{\label{tab:performance} Performance comparison between different ASICs suitable for SiPM time measurements.}
	\smallskip
	\begin{tabular}{|c|c|c|c|c|c|}
		\hline
		%\textbf{ASIC} & \textbf{Power in mW/channel} & \textbf{SPTR in ps FWHM} & \textbf{Input condition} & \textbf{Reference}\\
		\textbf{ASIC} & \textbf{Process} & \tabincell{c}{ \textbf{Power in}\\\textbf{mW/channel} } & \tabincell{c}{\textbf{SPTR in}\\\textbf{ps FWHM}}&\textbf{Tested SiPMs} & \textbf{Reference} \\		
		\hline
		NINO&250 nm&27&160& \tabincell{c}{S13360-3050CS \\ (10 V of VoV)  }&\cite{NINO,HRFlexToT}\\	
		\hline
		FlexToT&350 nm& 11 & 390 & \tabincell{c}{S13360-3050CS \\(4 V of VoV) } & \cite{FlexToT,HRFlexToT}\\
		\hline
		HRFlexToT&180 nm&3.5&260& \tabincell{c}{S13360-3050CS \\(4 V of VoV) } & \cite{HRFlexToT}\\
		\hline
		TOFPET2&110 nm&8.2&306& \tabincell{c}{S13361-3050AE-04 \\ (4 V of VoV) } & \cite{TOFPET2} \\
		\hline
		Petiroc2A&350 nm& 6 & 190 & \tabincell{c}{ S13360-3050CS \\ (7 V of VoV)} & \cite{Petiroc2A} \\
		\hline
		T11&180 nm& 4 & 250 & \tabincell{c}{S13360-3050PE \\ (4 V of VoV) } & This study \\
		\hline
		T12&180 nm& 3.8 & 254 & \tabincell{c}{S13360-3050PE \\ (4 V of VoV) } & This study \\
		\hline
		T22&180 nm& 3.8 & 264 & \tabincell{c}{S13360-3050PE \\ (4 V of VoV) } & This study \\
		\hline
		T1V&180 nm& 6.5 & 257 & \tabincell{c}{S13360-3050PE \\ (4 V of VoV) } & This study \\	
		\hline
	\end{tabular}
	%\vspace{1ex}
	%
	%{\raggedright $^{*}$T11 is the combination of CG and current discriminator I. T12 is the combination of CG and current discriminator II. T22 is the combination of NFBCG and current discriminator II. T1V is the combination of CG and voltage discriminator. \par}
\end{table}

\appendix
\section{Noise analysis of input stages}
\label{sec:noise}
Using the small signal model of the circuit on the left in figure~\ref{fig:input}, the current transfer function $H_{CG}$ can be obtained as equation \eqref{eq:a1}. 
\begin{equation}
	\label{eq:a1}
	\begin{split}
		H_{CG}(s)=\frac{I_{out}}{I_{in}}=-\frac{g_{m,MN1}g_{m,MP2}}{(sC_{x}+g_{m,MP1})(sC_{in}+g_{m,MN1})}.
	\end{split}
\end{equation}
Assuming an ideal impulse current input with charge $Q$, the output of the CG is illustrated in equation \eqref{eq:a2}. 
\begin{equation}
	\label{eq:a2}
	\begin{split}
		I_{out,CG}(t)=-\frac{g_{m,MN1}g_{m,MP2}Q}{g_{m,MN1}C_{x}-g_{m,MP1}C_{in}}(e^{-\frac{g_{m,MP1}}{C_{x}}t}-e^{-\frac{g_{m,MN1}}{C_{in}}t}).
	\end{split}
\end{equation}
The time constant of the falling edge is $\tau_{CG,falling}\approx\frac{C_{in}}{g_{m,MN1}}$. The maximum signal slope is at $t=0$ and can be approximated as equation \eqref{eq:a3} due to $C_{in} \gg C_{x}$ and $g_{m,MP2}=g_{m,MP1}$.
\begin{equation}
	\label{eq:a3}
	\begin{split}
		K_{CG,max}\approx\frac{g_{m,MN1}g_{m,MP1}Q}{C_{x}C_{in}}.
	\end{split}
\end{equation}
The noise $\sigma_{i}$ is mainly contributed by the input transistor MN1 and can be derived as equation \eqref{eq:a4}.
\begin{equation}
	\begin{aligned}
	\label{eq:a4}
		\sigma_{i,CG} &\approx\sqrt{\int_{0}^{+\infty}4\gamma kTg_{m,MN1}\left|\frac{g_{m,MP2}j2\pi fC_{in}}{(g_{m,MN1}+j2\pi fC_{in})(g_{m,MP1}+j2\pi fC_{x})}\right|^{2}df}
		\\
		 &= \sqrt{\frac{4\gamma kTg_{m,MN1}g_{m,MP2}^{2}C_{in}^{2}}{2\pi C_{in}^{2}C_{x}^{2}}\frac{\pi C_{in}C_{x}}{2(g_{m,MN1}C_{x}+g_{m,MP1}C_{in})}}
		 \\
		 &\approx\sqrt{\frac{\gamma kTg_{m,MN1}g_{m,MP1}}{C_{x}}}
	\end{aligned}
\end{equation}
where the factor $\gamma$ is $1/2$ for weak inversion, $k$ is the Boltzmann constant, $T$ is the temperature, and $g_{m,MP2}=g_{m,MP1}$. The time jitter of the CG can be computed as equation \eqref{eq:a5}.
\begin{equation}
	\label{eq:a5}
	\begin{split}
		\sigma_{t,CG}=\frac{\sigma_{i,CG}}{K_{CG,max}}\approx\frac{C_{in}\sqrt{\gamma kTC_{x}}}{Q\sqrt{g_{m,MN1}g_{m,MP1}}}.
	\end{split}
\end{equation}

The transfer function of NFBCG can be computed as
\begin{equation}
	\label{eq:a6}
	\begin{split}
		H_{NFBCG}(s)=\frac{-g_{m,MN1}g_{m,MP2}}{s^{2}C_{x}C_{in}+s(C_{x}g_{m,MN1}+C_{in}g_{m,MP1})+g_{m,MN1}(g_{m,MP1}+g_{m,MP3})}.
	\end{split}
\end{equation}
The output of the NFBCG is illustrated in equation \eqref{eq:a7} with the aforementioned input.
\begin{equation}
	\label{eq:a7}
	\begin{split}
		I_{out,NFBCG}(t)\approx-\frac{g_{m,MN1}g_{m,MP2}Q}{(g_{m,MN1}+2g_{m,MP3})C_{x}-g_{m,MP1}C_{in}}(e^{s_{1}t}-e^{s_{2}t})
	\end{split}
\end{equation}
where $s_{1}=-\frac{g_{m,MP1}}{C_{x}}+\frac{g_{m,MN1}g_{m,MP3}}{C_{in}g_{m,MP1}}$ and $s_{2}=-\frac{g_{m,MN1}(g_{m,MP1}+g_{m,MP3})}{C_{in}g_{m,MP1}}$. The time constant of the falling edge is $\tau_{NFBCG,falling}\approx\frac{C_{in}g_{m,MP1}}{g_{m,MN1}(g_{m,MP1}+g_{m,MP3})}$. Similarly, the maximum signal slope is at $t=0$ and can be approximated as equation \eqref{eq:a8}, which is almost the same as equation \eqref{eq:a3}.
\begin{equation}
	\label{eq:a8}
	\begin{aligned}
		K_{NFBCG,max}&\approx\frac{g_{m,MN1}g_{m,MP2}Q}{g_{m,MP1}C_{in}-g_{m,MN1}(1+\frac{2g_{m,MP3}}{g_{m,MP1}})C_{x}}\left(\frac{g_{m,MP1}}{C_{x}}-\frac{g_{m,MN1}g_{m,MP3}}{C_{in}g_{m,MP1}}\right)
		\\
		&\approx\frac{g_{m,MN1}g_{m,MP2}Q}{C_{x}C_{in}}.
	\end{aligned}
\end{equation}
Equation \eqref{eq:a9} can be obtained by considering only the noise of input transistor MN1.
\begin{equation}
	\begin{aligned}
		\label{eq:a9}
		\sigma_{i,NFBCG}^{2} &\approx\int_{0}^{+\infty}4\gamma kTg_{m,MN1}\times\\
			&\left|\frac{g_{m,MP2}j2\pi fC_{in}}{(g_{m,MN1}+j2\pi fC_{in})(g_{m,MP1}+j2\pi fC_{x})+g_{m,MN1}g_{m,MP3}}\right|^{2}df
		\\
		&\approx\int_{0}^{+\infty}4\gamma kTg_{m,MN1}\left|\frac{g_{m,MP2}j2\pi fC_{in}}{C_{in}C_{x}(j2\pi f-s_{1})(j2\pi f-s_{2})}\right|^{2}df
		\\
		&=\frac{4\gamma kTg_{m,MN1}g_{m,MP2}^{2}}{2\pi C_{x}^{2}}\frac{\pi}{2(|s_{1}|+|s_{2}|)}
		\\
		&\approx\frac{\gamma kTg_{m,MN1}g_{m,MP2}^{2}}{C_{x}g_{m,MP1}}.
	\end{aligned}
\end{equation}
The time jitter of the NFBCG can be computed as equation \eqref{eq:a10} which is the same as equation \eqref{eq:a5}.
\begin{equation}
	\label{eq:a10}
	\begin{split}
		\sigma_{t,NFBCG}\approx\frac{C_{in}\sqrt{\gamma kTC_{x}}}{Q\sqrt{g_{m,MN1}g_{m,MP1}}}.
	\end{split}
\end{equation}

\acknowledgments

This work was supported by the Ministry of Science and Technology of China (No. 2022YFA1605500), Shanghai Pilot Program for Basic Research — Shanghai Jiao Tong University (No. 21TQ1400218), and Yangyang Development Fund. The authors would like to thank Jun Guo, Hualin Mei and Yong Yang for their help to improve this paper.



\begin{thebibliography}{99}
\bibitem{Trident} Ye Z.P. et al., \emph{A multi-cubic-kilometre neutrino telescope in the western Pacific Ocean}, \emph{Nature Astronomy} {\bf 7} (2023) 1497.
\bibitem{TridentE} M. X. Wang et al., \emph{Design of the Readout Electronics for the TRIDENT Pathfinder Experiment}, \emph{IEEE Transactions on Nuclear Science} {\bf 70} (2023) 2240.
\bibitem{hDOM} F. Hu, Z. Li and D. Xu, \emph{Exploring a PMT+SiPM hybrid optical module for next generation neutrino telescopes}, \emph{PoS ICRC2021} (2021) 1043.
\bibitem{sipmdcr} Adam Nepomuk Otte et al., \emph{Characterization of three high efficiency and blue sensitive silicon photomultipliers}, \emph{Nucl. Instrum. Methods Phys. Res. A, Accelerators Spectrometers Detectors Assoc. Equip.}  {\bf 846} (2017) 106.
\bibitem{sipm1} W. Zhi et al., \emph{Preliminary Design of the Hybrid Digital Optical Module for TRIDENT}, \emph{PoS ICRC2023} (2023) 1213.
\bibitem{sipm2} W. Zhi et al., \emph{Front-end electronics development of large-area SiPM arrays for high-precision single-photon time measurement}, \emph{J. Instrum.} {\bf 19} (2024) P06011.
\bibitem{NINO} F. Anghinolfi et al., \emph{NINO: An ultra-fast and low-power frontend amplifier/discriminator ASIC designed for the multigap resistive plate chamber}, \emph{Nucl. Instrum. Methods Phys. Res. A, Accelerators Spectrometers Detectors Assoc. Equip.}  {\bf 533} (2004) 183. 
\bibitem{FlexToT} I. Sarasola et al., \emph{A comparative study of the time performance between NINO and FlexToT ASICs}, \emph{J. Instrum.} {\bf 12} (2017) P04016.
\bibitem{HRFlexToT} David Sánchez et al., \emph{HRFlexToT: A High Dynamic Range ASIC for Time-of-Flight Positron Emission Tomography}, \emph{IEEE Transactions on Nuclear Science} {\bf 6} (2022) 51.
\bibitem{Petiroc2A} S. Ahmad et al., \emph{Petiroc2A: Characterization and experimental results}, \emph{Proc. IEEE Nucl. Sci. Symp. Med. Imag. Conf. (NSS/MIC)} Sydney, NSW, Australia, Nov. 2018, pp. 1.
\bibitem{TOFPET2} R. Bugalho et al., \emph{Experimental characterization of the TOFPET2 ASIC}, \emph{J. Instrum.} {\bf 14} (2019) P03029.
\bibitem{s13360} Hamamatsu Photonics K.K., \emph{MPPC S13360 series}, https://www.hamamatsu.com/content/dam/hamamatsu-photonics/sites/documents/99\_SALES\_LIBRARY/ssd/s13360\_series\_kapd1052e.pdf.
\bibitem{DIET} Zijian Lang, Z. Deng and H. Gong, \emph{Evaluation of the timing performance of the DIET ASIC}, \emph{J. Instrum.} {\bf 18} (2023) P03010.
\bibitem{disc1} Hongchin Lin, Jie-Hau Huang and Shyh-Chyi Wong, \emph{A simple high-speed low current comparator}, \emph{IEEE International Symposium on Circuits and Systems (ISCAS)} {\bf 2} (2000) 713.
\bibitem{disc2} Chen L., Shi B. Lu C., \emph{Circuit Design of a High Speed and Low Power CMOS Continuous-time Current Comparator}, \emph{Analog Integrated Circuits and Signal Processing}  {\bf 28} (2001) 293.
\bibitem{summation} R. Manera et al., \emph{Optimizing time resolution and power consumption in a current-mode circuit for SiPMs}, \emph{J. Instrum.} {\bf 19} (2024) T04009.
\bibitem{MUSIC} Sergio Gómez et al., \emph{MUSIC: An 8 channel readout ASIC for SiPM arrays}, \emph{Proc. of SPIE} {\bf 9899} (2016) Optical Sensing and Detection IV, 98990G.
\bibitem{EXYT} Xuezhou Zhu, Zhi Deng, Yu Chen, Yinong Liu, and Yaqiang Liu, \emph{Development of a 64-Channel Readout ASIC
for an 8 $\times$ 8 SSPM Array for PET and TOF-PET Applications}, \emph{IEEE Transactions on Nuclear Science} {\bf 63} (2016) 1327.
\bibitem{DIET2} Y. Chen, Z. Deng and Y. Liu, \emph{DIET: a multi-channel SiPM readout ASIC for TOF-PET with individual energy and timing digitizer}, \emph{J. Instrum.} {\bf 13} (2018) P07023.
\bibitem{sipmmodel} A. Di Francesco et al.,  \emph{TOFPET2: a high-performance ASIC for time and amplitude measurements of SiPM signals in time-of-flight applications}, \emph{J. Instrum.}, {\bf 11} (2016) C03042.


\end{thebibliography}
\end{document}